\documentclass[preprint]{ptephy_v1}

\preprintnumber{XXXX-XXXX} 
\usepackage{hyperref}
\usepackage{ulem}
\usepackage{color}

\newcommand{\void}[1]{}
\renewcommand{\color}{\void}

\begin{document}

\title{Norm operator method for boson expansions}


\author{Kimikazu Taniguchi}
\affil{Department of Medical Information Science, Suzuka University of Medical Science, Suzuka  1001-1, Japan\email{kimikazu@suzuka-u.ac.jp}}





\begin{abstract}
We propose a new boson expansion theory \color{blue} that does not premise \color{black} the closed-algebra approximation, indispensable for formulation until now, as an extension of the conventional practical boson expansion methods that have tried to elucidate nuclear collective motion,  \color{blue} which is a method that allows the closed-algebra approximation not to be used or to be used appropriately, enables us to obtain the boson expansion easier, and reproduces the fermion subspace onto the boson subspace more faithfully than the conventional practical methods.
The two-phonon norm matrix composed of all the phonon excitation modes is investigated in detail, which reveals the mechanism, essential for the boson expansion methods, of how we should construct the fermion subspace to be mapped from the whole fermion space.
The conventional practical boson expansion methods have applied the closed-algebra approximation improperly to strengthen the effect of the Pauli principle inappropriately,
which should be replaced by those that do not use that approximation or use it properly.
\color{black}
\end{abstract}

\subjectindex{xxxx, xxx}

\maketitle

\section{Introduction}
Microscopic elucidation of the large-amplitude collective motion of atomic nuclei remains one of the important and challenging tasks, and its achievement requires developing a method to overcome small-amplitude-oscillation approximations like the Tamm-Dancoff approximation and the random phase approximation. The boson expansion theory is one of the methods going beyond the small-amplitude-oscillation approximations \cite{KM91}. 

The initial formulation of the boson expansion theory was given by replacing fermion quasiparticle pair operators with boson polynomials that reproduce their commutation relations \cite{BZ62}.
Later, referring to the preceding work \cite{Usi60}, the boson expansion theory has been given as a mapping theory by utilizing a one-to-one correspondence between the basis vector in fermion space and the completely antisymmetric state vector in boson space \cite{MYT64}. The Holstein-Primakoff and the Dyson Boson expansions also have been formulated in the same way \cite {JDF71}. These formulations target all pair operator excitations in fermion space.

Practical use, however, has not adopted the total excitation modes but the collective and, in case of need, some non-collective excitation modes of phonons \cite{LH75, KT76, HJJ76}.
One is the method to pick up only the phonons regarded as essential and seek boson expansions that reproduce their commutation relations \cite {KT76, KT72, KT83}, and the other to construct the mapping operator with only the excitation modes regarded as essential \cite{LH75, HJJ76}.
However, the boson expansions derived from the former method received the claim that it was chimerical \cite{Ma80a, Ma80b}.
For refutation, it has been rederived as a normal-ordered linked-cluster expansion (NOLCEXP)  using the latter method \cite{KT83, SK88}.
Later, the Dyson boson expansion theory (DBET), finite expansion of non-Hermitian type, has also adopted the latter method for formulation \cite {Ta01}. 
The latter method has also given reason for the use of the ideal boson state vectors, which do not consider the effects of the Pauli exclusion principle at all \cite {LH75, KT83}. These boson expansion methods have contributed to the elucidation of the large-amplitude collective motions such as the shape transition of nuclei in the transitional region \cite{KT76, TTT87, SK91}. 
In addition, there is another attempt to derive NOLCEXP from dynamical nuclear field theory \cite {KKS91}.

Despite such contributions of the boson expansion theory, there are problems to be solved.

One is whether the closed-algebra approximation is \color{blue}applied appropriately \color{black} or not.
 Fermion subspace is constructed based on multi-phonon state vectors consisting of collective excitation modes and some non-collective ones of the Tam-Dancoff type phonons. The commutation relations among the Tamm-Dancoff phonons do not close among the so-selected ones. 
Until now without exception, the boson expansion methods with restriction of the phonon excitation modes have used approximations that neglect the unselected modes.
NOLCEXP neglects these in the inverse of the norm matrices to perform boson expansions and suppresses all remaining ones in the obtained boson expansions \cite{KT83}.
DBET adopts the approximation named phonon-truncation approximation \cite{TTT87}, also called {\it closed-algebra approximation} \cite{Ta01}, which truncates the unselected phonon operators.
Each approximation above is essential for NOLCEXP and DBET: NOLCEXP adopts it for obtaining the finite expansions for the scattering operators necessary to refute the claim that it is chimerical, and DBET to obtain the phonon and scattering operators to be finite expansions.
All these prescriptions bring the closed-algebra approximation, which closes the commutation relations among the selected phonon operators.
The Tamm-Dancoff approximation \cite{TTT87} or similar approximations \cite{KT76, SK91} determine the excitation modes of the Tamm-Dancoff type phonons. They are small-amplitude-oscillation approximations, and therefore only the collective excitation modes selected by these methods are not sufficient for reproducing the shape transition of the nucleus in the transitional region. Applications of the boson expansions adopting the closed-algebra approximation have revealed the necessity of the contribution of some non-collective excitation modes \cite{KT76, TTT87}. The only way to introduce the contribution of the non-collective excitation modes \color{blue} by the conventional practical boson expansion methods \color{black} is to introduce bosons having these modes and perform the boson expansions.
It is known that the boson expansions for only the collective modes have good convergence \cite{KT76, KT83}. Those for the non-collective excitation modes, however, have raised the question of whether the boson expansions for the non-collective excitation modes converge well and led to claim of the superiority of DBET as finite expansions \cite{TTT87}. However, we have a dispute about how to deal with the non-Hermite expansions \cite{KTM99, SST99}. On the other hand, without the closed-algebra approximation \color{blue} or with proper use of it\color{black}, we would be able to take the contribution of the non-collective modes into the coefficients of the boson expansions consisting of only the bosons with collective modes and avoid these problems.
\color{blue}
We should formulate the boson expansion method which does not premise the closed-algebra approximation but can adopt it according to necessity.
It enables us to investigate whether the closed-algebra approximation is applied appropriately and also makes it possible to include the contribution of the non-collective excitation modes without the convergence problem.
\color{black}

The other is about the phonon excitation number. Until now, for the multi-phonon state vectors, used as the basis vectors of the fermion subspace to be mapped, the sorts of the excitation modes have been limited, while the number of phonon excitations has not \cite{KT83, SK88, Ta01}. No restriction of the phonon excitation number permits the eigenvalues of the multi-phonon state vector to become zero as the number of excitations increases. 
In DEBT, the mapped operators become the product of the finite boson expansion part and the projection operator onto the physical subspace, which is the boson subspace corresponding to the preselected fermion subspace.  Reflecting the effect of the Pauli exclusion principle, the projection operator is generally very complex. For small excitation numbers, however, the multi-phonon norm matrix does not have zero eigenvalues, and we can replace the projection operator with the unit operator of the boson space, which enables us to obtain finite expansions.
Moreover, its formulation indicates that we can formulate DEBT using a mapping operator limiting the number of phonons. It also indicates that both limiting the number of bosons after the mapping and doing that before mapping results in the same.
These results are, however, obtained by using the closed-algebra approximation, and it is not clear whether they hold without it.
NOLCEXP uses the boson expansion of the projection operator onto the vacuum of the boson space, whose convergence is not guaranteed, and the expansions of the norm matrices of the multi-phonon state vectors. NOLCEXP can not be achieved without no limitation of the phonon excitation number and an unnatural assumption that the norm matrices have no zero eigenvalues in any case \cite{KT83}. Hence the validity of limiting the number of boson excitations after mapping is not self-evident.
We should formulate the boson expansion theory with the number of phonon excitations limited beforehand.

In addition, NOLCEXP has obtained only the terms up to the next-to-leading order of magnitude for the phonon operators
\color{blue}
and the scattering operators \cite{SK88}.
\color{black}
NOLCEXP \cite{KT83} has been utilized to justify the boson expansion method \cite{KT76} applied in realistic cases formulated by commutation relations.
Initially, NOLCEXP derived the leading-order terms of the scattering operators, however, it was the previous method \cite{KT76} based on the commutation relations that derived the higher-order expansion coefficients for the phonon operators.
It is another paper published later that has derived coefficients from NOCEXP \cite{SK88}. NOLCEXP requires {\it elementary but rather tedious manipulations}.
We should develop a technique that can derive higher-order terms more easily not by the method using commutation relations but by the method using a mapping operator.

We have proposed a boson-fermion expansion theory \cite{TM90, TM91} as an alternative to the boson expansion theory.
Whereas the boson expansion theory treats all the adopted phonon excitation modes as bosons, this method treats, reflecting the differences in the properties of these excitation modes, in zeroth-order approximation, all the collective excitation modes as bosons and all the non-collective ones as fermions. 
From this method, we can derive the boson expansions, which should be compared, by extending the boson part up to the non-collective modes necessary and depressing the fermion excitations. 
We have shown that by adopting an appropriate transformation operator in boson-fermion space, the derived expansions become the same as those of NOLCEXP \cite{TKM94}. Since the formulation of the boson-fermion expansion theory has not used the closed-algebra approximation, it would be worthwhile to formulate a new boson expansion method that does not use the closed-algebra approximation and compare it with this method.

In this paper, we propose a boson expansion method that becomes an extension of the conventional practical boson expansion method \color{blue}
which does not premise the closed-algebra approximation but can adopt it according to necessity as well as makes it possible to check whether the ideal boson state vectors have no spurious components. 
\color{black}
We give a mapping operator that limits both the number of phonon excitations and the sorts of phonon excitation modes to perform proper boson expansions which bring both the Hermitian and the non-Hermitian types of boson expansions. The norm operator makes it possible to perform boson expansions easier. We give the higher-order terms not obtained so far by the mapping methods and the necessary condition for boson expansions to be small parameter expansions. 
Without the closed-algebra approximation
\color{blue} or with the proper use of it, 
\color{black} boson expansions become infinite regardless of whether the mapping is the Hermitian or the non-Hermitian type. 
The closed-algebra approximation
\color{blue}
that closes the commutation relations among the phonons to be boson expanded
holds only when the small parameter expansions become impossible as well as the ideal boson state vectors include the spurious
components.
The conventional boson expansion methods apply the closed-algebra approximation to the small parameter expansions actually, which is improper and causes the neglection of the important terms of boson expansions, while the appropriate use of the closed-algebra approximation enables us to remove the non-effective phonon excitation modes properly.
\color{black}
No use of the closed-algebra approximation
\color{blue} or the proper use of it
\color{black} enables us to incorporate the contribution of all
\color{blue} or enough
\color{black}
sorts of non-collective excitation modes into the coefficients of the boson expansions consisting of only bosons with collective modes with avoiding the convergence problem of the boson expansions for non-collective modes.
\color{blue}
The investigation of the closed-algebra approximation using the two-phonon norm matrix reveals the mechanism of the convergence of the boson expansions, which has been mistaken \cite{HJJ76} or not fully understood \cite{KT83, SK88}.
\color{black}

Section 2 deals with the Tam-Dancoff type phonon, the multi-phonon state vector, and the ideal boson state vector.

Section 3 deals with the mapping. We give a mapping operator that limits both the number of phonon excitations and the sorts of phonon excitation modes to perform proper boson expansions. Next, we write the mapping operator as the product of the power of the norm operator and the part common to all types of mapping and clarify the difference in mapping types. We also show that the boson mapping results in the boson approximation when the maximum phonon excitation number is 1, which indicates that the boson approximation has been established as the boson mapping whose maximum phonon excitation number is 1.

Section 4 deals with boson expansions. 
\color{blue}
We investigate the two-phonon norm matrix composed of all the phonon excitation modes in detail and reveal the mechanism, which is essential for the boson expansion methods, of how we should construct the fermion subspace to be mapped from the whole fermion space.
\color{black}
We give the necessary condition where the small parameter expansions hold and present how to evaluate the order of magnitude of the boson expansion terms.
We investigate and discuss the approximations that neglect the unselected phonon excitation modes, that is, phonon-truncation approximation, etc., which the conventional practical boson expansion methods adopt and bring the closed-algebra approximation.
\color{blue}
We also investigate the closed-algebra approximation itself and present the possibility of its application that the conventional boson expansion methods have not adopted.
\color{black}
We give formulas for the boson expansions without the closed-algebra approximation and obtain the expansion terms neglected under the closed-algebra approximation as well as the higher-order terms not obtained until now by the mapping methods. 
\color{blue}
It also allows us to use the closed-algebra approximation properly.
We also give the boson expansion of the norm operator, which not only facilitates the boson expansions but also makes it possible to confirm that the mapping is performed within the range of the appropriate phonon excitation number.
\color{black}
We propose a new method to incorporate the contribution of all
\color{blue}
or enough
\color{black}
non-collective modes without annoying us with the convergence problem of the boson expansions for them.
We also discuss the boson expansions under the situation that the closed-algebra approximation \color{blue} makes the small parameter expansion impossible\color{black}.

Section 5 refers to the comparison to the boson-fermion expansion theory.

Section 6 summarizes.

\section{Fermion space and boson space}

\subsection{Tamm-Dancoff type phonon operators, scattering operators, and their commutation relations
}
The definition of the Tam Dancoff (TD) type phonon creation and annihilation operators are as follows:
\begin{subequations}
\label{eq:phononop}
\begin{equation}
\label{eq:phononopc}
X_\mu^\dagger
=\displaystyle\sum_{\alpha<\beta}\psi_\mu(\alpha\beta)a_\alpha^\dagger
a_\beta^\dagger,
\end{equation}
\begin{equation}
\label{eq:phononopa}
X_\mu=\displaystyle\sum_{\alpha<\beta}\psi_\mu(\alpha\beta)a_\beta
a_\alpha.
\end{equation}
\end{subequations}
Here, $a_\alpha^\dagger$ and $a_\alpha$ are quasi-particle creation and annihilation operators in a one-particle state $\alpha$.
The coefficients satisfy the following relations:
\begin{subequations}
\label{eq:TDrel}
\begin{equation}
\label{eq:anti}
\psi_\mu(\beta\alpha)=-\psi_\mu(\alpha\beta)
\end{equation}
\begin{equation}
\label{eq:orthonormal}
\sum_{\alpha<\beta}\psi_\mu(\alpha\beta)\psi_{\mu'}(\alpha\beta)=\delta_{\mu,
\mu'},
\end{equation}
\begin{equation}
\label{eq:complete}
\sum_{\mu}\psi_\mu(\alpha\beta)\psi_{\mu}(\alpha'\beta')=\delta_{\alpha,
\alpha'}\delta_{\beta, \beta'}-\delta_{\alpha,
\beta'}\delta_{\beta, \alpha'}.
\end{equation}
\end{subequations}

Next, we introduce the scattering operators,
\begin{equation}
\label{eq:scop}
B_q=\sum_{\alpha\beta}\varphi_q(\alpha\beta)a_\beta^\dagger
a_\alpha,
\end{equation}
where the coefficients $\varphi_q(\alpha\beta)$ satisfy
\begin{subequations}
\label{eq:coeffscop}
\begin{equation}
\label{eq:coeffscop1}
\displaystyle\sum_{\alpha\beta}\varphi_q(\alpha\beta)\varphi_{q'}(\alpha\beta)=\delta_{q,
q'},
\end{equation}

\begin{equation}
\label{eq:coeffscop2}
\displaystyle\sum_{q}\varphi_q(\alpha\beta)\varphi_{q}(\alpha'\beta')=\delta_{\alpha,\alpha'}\delta_{\beta,
\beta'}.
\end{equation}
\end{subequations}

In addition, we introduce $\bar q$ such that
\begin{equation}
\label{eq:barq}
\varphi_{\bar q}(\alpha\beta)=\varphi_q(\beta\alpha).
\end{equation}
By this definition, $B_{\bar q}=B_q^\dagger$.

The phonon and scattering operators satisfy the following commutation relations:
\begin{subequations}
\label{eq:algebra}
\begin{equation}
\label{eq:algebra1}
[ X_\mu, X_{\mu'}^\dagger ]=\delta_{\mu,
\mu'}-\sum_q\Gamma^{\mu\mu'}_qB_q,
\end{equation}

\begin{equation}
\label{eq:algebra2}
[ B_q, X_\mu^\dagger
]=\sum_{\mu'}\Gamma^{\mu\mu'}_qX_{\mu'}^\dagger,
\end{equation}

\begin{equation}
\label{eq:algebra3}
[ X_\mu, B_q ]=\sum_{\mu'}\Gamma^{\mu'\mu}_qX_{\mu'},
\end{equation}
\end{subequations}
where the definition of $\Gamma^{\mu\mu'}_q$ is as follows:
\begin{equation}
\label{eq:Gamma}
\Gamma^{\mu\mu'}_q=\sum_{\alpha\beta}\varphi_q(\alpha\beta)\Gamma^{\mu\mu'}_{\alpha\beta},\quad
\Gamma^{\mu\mu'}_{\alpha\beta}=\sum_\gamma\psi_\mu(\alpha\gamma)\psi_{\mu'}(\beta\gamma).
\end{equation}
The following relation holds:
\begin{equation}
\label{eq:qbargam}
\Gamma_{\bar q}^{\mu_1 \mu_2}=\Gamma_q^{\mu_2 \mu_1}.
\end{equation}

From Eqs. (\ref{eq:algebra1}) and (\ref{eq:algebra2}), we obtain
\begin{equation}
\label{eq:doublecom}
[ [X_{\mu_1}, X_{\mu_2}^\dagger], X_{\mu_3}^\dagger] = -\sum_{\mu'}Y(\mu_1, \mu_2, \mu_3, \mu')X_{\mu'}^\dagger,
\end{equation}
where the definition of $Y(\mu_1\mu_2\mu_3\mu_4)$ is 
\begin{equation}
\label{eq:Y}
Y(\mu_1\mu_2\mu_3\mu_4)=\sum_q\Gamma_q^{\mu_1\mu_2}\Gamma_q^{\mu_3\mu_4}
=\sum_{\alpha\beta}\Gamma_{\alpha\beta}^{\mu_1\mu_2}\Gamma_{\alpha\beta}^{\mu_3\mu_4}.
\end{equation}
The following relation holds:
\begin{equation}
\label{eq:Ysym}
\begin{array}{lll}
Y(\mu_1\mu'_1\mu'_2\mu_2)
&=&Y(\mu_2\mu'_1\mu'_2\mu_1)
\\
&=&Y(\mu_1\mu'_2\mu'_1\mu_2)
\\
&=&Y(\mu'_1\mu_1\mu_2\mu'_2).
\\
\end{array}
\end{equation}

\subsection{Multi-phonon and multi-boson state vectors}
We divide the phonon excitation modes $\{\mu\}$ into two groups, $\{t\}$ and $\{\bar t\}$,and prepare the multi-phonon state vectors, 
\begin{equation}
\label{eq:mulphst}
\vert N; t\rangle\rangle\rangle=\vert t_1, t_2, \cdots ,
t_N\rangle\rangle\rangle=X_{t_1}^\dagger X_{t_2}^\dagger \cdots
X_{t_N}^\dagger \vert 0\rangle\quad (0\leq N\leq N_{max}),
\end{equation}
which are not generally normalized and not orthogonal to one another. Here $\vert 0\rangle$ is the vacuum of quasi-particles.

For obtaining orthonormal basis vectors for the fermion subspace to be mapped onto a boson subspace, we first introduce state vectors, 
\begin{equation}
\label{eq:bnmulphst}
\vert N; t\rangle=\mathcal{N}_B(N; t)^{-1/2}\vert N;
t\rangle\rangle\rangle,
\end{equation}
where $\mathcal{N}_B(N; t)$ are the normalization factors that normalize the multi-phonon state vectors under the boson approximation $[X_t, X_{t'}^\dagger]\approx \delta_{t, t'}$, and solve the eigen equations for the norm matrices of the multi-phonon state vectors,
\begin{equation}
\label{eq:eigeneqnormmtx}
\sum_{t'}\langle N; t\vert N; t'\rangle u_a^{t'}(N)=z_a(N)u_a^t(N).
\end{equation}
The eigenvalues $z_a(N)$ become positive or zero. Hereafter, $a_0$ means that its eigenvalue is zero. 
The eigenvectors are orthonomalized as
\begin{subequations}
\label{eq:eigenvcon}
\begin{equation}
\label{eq:eigenvcon1}
\displaystyle\sum_tu_a^t(N)u_{a'}^t(N)=\delta_{a, a'},
\end{equation}
and satisfy the completeness relations
\begin{equation}
\label{eq:eigenvcon2}
\displaystyle\sum_au_a^t(N)u_a^{t'}(N)=\delta_{t, t'}.
\end{equation}
\end{subequations}
The orthonormalized basis vectors are given by
\begin{equation}
\label{eq:orthonormbs}
\vert N; a\rangle\rangle = z_a^{-\frac 12}(N)\sum_tu_a^t(N)\vert N; t\rangle
\qquad (a\neq a_0).
\end{equation}

The projection operator onto the fermion subspace spanned by the basis vectors of Eq. (\ref{eq:mulphst}) or Eq. (\ref{eq:orthonormbs}) is given by
\begin{equation}
\label{eq:unitf}
\hat T_F=\sum_{N=0}^{N_{max}}\hat T_F(N),
\end{equation}
where
\begin{equation}
\label{eq:proptfn}
\begin{array}{l}
\hat T_F(N)=\displaystyle\sum_{a\neq a_0}\vert N; a\rangle\rangle\langle\langle N; a\vert \quad (N\ge 1),
\\
\hat T_F(0)=\vert 0\rangle\langle 0\vert.
\end{array}
\end{equation}
The projection operator onto the fermion space composed of even quasi-particles becomes
\begin{equation}
\begin{array}{lll}
\hat 1_F&=&\displaystyle\lim_{N\rightarrow\infty}\sum_{\alpha_1<\beta_1<\cdots <\alpha_N<\beta_N}\vert\alpha_1\beta_1\cdots\alpha_N\beta_N\rangle\langle\alpha_1\beta_1\cdots\alpha_N\beta_N\vert
\\
&=&\displaystyle\lim_{N_{max}\rightarrow\infty}\hat T_F,
\end{array}
\end{equation}
where $\vert\alpha_1\beta_1\cdots\alpha_N\beta_N\rangle=a_{\alpha_1}^\dagger a_{\beta_1}^\dagger\cdots a_{\alpha_N}^\dagger a_{\beta_N}^\dagger \vert 0\rangle$.
Note that,
\begin{equation}
\label{eq:multpha0}
\vert N; a_0\rangle\rangle=\sum_{t}u_{a_0}^t(N)\vert N; t\rangle
\end{equation}
satisfies
\begin{equation}
\label{eq:nultpha0norm}
\langle\langle N; a_0\vert N; a_0\rangle\rangle=0,
\end{equation}
therefore $\vert N; a_0\rangle\rangle=0$.

Next we introduce boson creation and annihilation operators, $b_t^\dagger$ and $b_{t'}$, having the same indices as those of the multi-phonons, $X_t^\dagger$ and $X_{t'}$:
\begin{equation}
\label{eq:bcom}
[ b_t, b_{t'}^\dagger ] = \delta_{t, t'}.
\end{equation}
The multi-boson states,
\begin{equation}
\label{eq:multib}
\vert N; t)))=\vert t_1, t_2, \cdots , t_N)))=b_{t_1}^\dagger
b_{t_2}^\dagger \cdots b_{t_N}^\dagger \vert 0),
\end{equation}
are orthogonal to one another, and are normalized by their norms,
\begin{equation}
\label{eq:normb}
\mathcal{N}_B(N; t)=(((N: t\vert N; t))),
\end{equation}
such as
\begin{equation}
\label{eq:bbasis}
\vert N; t)=\vert t_1, t_2, \cdots , t_N)=\mathcal{N}_B(N; t)^{-1/2}\vert N; t))).
\end{equation}
They are so-called ideal boson state vectors.

The projection operator onto the boson subspace  where the maximum phonon excitation number is $ N_ {max} $ is given by
\begin{equation}
\label{eq:brave1b}
\breve 1_B=\sum_{N=0}^{N_{max}}\hat 1_B(N).
\end{equation}
$\hat 1_B(N)$ is  the projection operator onto the boson subspace where the boson excitation number is $N$:
\begin{equation}
\label{eq:1bn}
\begin{array}{lll}
\hat 1_B(N)&=&\displaystyle\sum_{t}\vert N; t)(N; t\vert
\\
&=&\displaystyle\sum_{t_1\le\cdots\le t_N}\vert t_1\cdots t_N)(t_1\cdots t_N\vert
\\
&=&\displaystyle\sum_{t_1\cdots t_N}\frac 1{N!}\vert t_1\cdots t_N)))(((t_1\cdots t_N\vert\qquad (N\ge 1),
\\
\hat 1_B(0)&=&\vert 0)(0\vert.
\end{array}
\end{equation}
Here we use  that the function $f(t_1, \cdots, t_N)$, which is completely symmetric to the argument, satisfies the following \cite{KKS91}:
\begin{equation}
\sum_{t_1\leq\cdots\leq t_N}f(t_1, \cdots , t_N)=\sum_{t_1, \cdots , t_N}\frac{\mathcal{N}_B(t_1, \cdots , t_N)}{N!}f(t_1, \cdots ,t_N).
\end{equation}

The unit operator of the boson space is
\begin{equation}
\label{eq:bunitop}
\hat 1_B=\lim_{N_{max}\rightarrow\infty}\breve 1_B.
\end{equation}

The following state vectors
\begin{equation}
\label{eq:bbsistr}
\vert N; a))=\sum_tu_a^t(N)\vert N; t)
\end{equation}
also become orthonormal basis vectors.
From Eqs. (\ref{eq:eigenvcon1}) and  (\ref{eq:eigenvcon2}),
\begin{subequations}
\label{eq:nabborthonormcomplb}
\begin{equation}
\label{eq:nabborthonormb}
((N; a\vert N'; a'))=\delta_{N, N'}\delta_{a, a'},
\end{equation}
\begin{equation}
\label{eq:nabbcomplb}
\sum_a \vert N; a))((N; a\vert=\hat 1_B(N).
\end{equation}
\end{subequations}
Introducing projection operators:
\begin{equation}
\hat T_B(N)=\sum_{a\neq a_0}\vert N; a))((N; a\vert,\quad \hat T^{(0)}_B(N)=\sum_{a_0}\vert N:a_0))((N; a_0\vert,
\end{equation}
\begin{equation}
\label{eq:proptb}
\hat T_B=\sum_{N=0}^{N_{max}}\hat T_B(N),\quad\hat T^{(0)}_B=\sum_{N=0}^{N_{max}}\hat T^{(0)}_B(N),
\end{equation}
following relations hold:
\begin{subequations}
\label{eq:idealwith0}
\begin{equation}
\hat 1_B(N)=\hat T_B(N)+\hat T^{(0)}_B(N),
\end{equation}
\begin{equation}
\breve 1_B=\hat T_B+\hat T^{(0)}_B.
\end{equation}
\end{subequations}
Especially when the phonon excitation mode and the maximum number of excitations are selected so that the norm matrix has no zero eigenvalues, then 
\begin{subequations}
\label{eq:idealno0}
\begin{equation}
\hat 1_B(N)=\hat T_B(N)\quad (N\leq N_{max}),
\end{equation}
\begin{equation}
\breve 1_B=\hat T_B,
\end{equation}
\end{subequations}
hold.

\section{Boson mapping}
This section deals with boson mapping. We introduce a mapping operator from fermion subspace to boson subspace that limits the number of phonon excitations, which the conventional boson expansion methods have not adopted, in addition to the phonon excitation mode of the multi-phonon state vectors. The norm operator in the mapping operator clarifies the relationship between the Hermitian type and the non-Hermitian types and makes boson expansions obtained more easily.  
\subsection{Mapping operator}
The fermion state vector, $\vert N; a\rangle\rangle$ and the boson state vector $\vert N; a))$ have a one-to-one correspondence with each other when $a\neq a_0$, which we utilize to make a mapping operator, 
\begin{equation}
\label{eq:bmop}
U_\xi=\sum_{N=0}^{N_{max}}U_\xi(N);\quad U_\xi(N)=\sum_{a\neq a_0}z_a(N)^\xi\vert N; a))\langle\langle N; a\vert.
\end{equation}
The following relations are satisfied:
\begin{equation}
\label{eq:utftbh}
U_{-\xi}^\dagger U_{\xi}=\hat T_F,\qquad U_{\xi}U_{-\xi}^\dagger =\hat T_B.
\end{equation}

This operator maps the state vectors and operators of fermion space onto those of boson subspace:
\begin{subequations}
\label{eq:ximap}
\begin{equation}
\label{eq:ximap1}
\vert \psi')_{\xi}= U_{\xi}\vert\psi'\rangle,\qquad {}_{-\xi} (\psi\vert =\langle\psi\vert U_{-\xi}^\dagger,
\end{equation}
\begin{equation}
\label{eq:ximap2}
(O_F)_{\xi}=U_{\xi}O_FU_{-\xi}^\dagger.
\end{equation}
\end{subequations}
These satisfy the following relations:
\begin{subequations}
\label{eq:ximappm}
\begin{equation}
\label{eq:ximappm1}
\vert \psi')_{\xi}=\left\{{}_\xi(\psi'\vert\right\}^\dagger,\qquad {}_{-\xi} (\psi\vert =\left\{\vert\psi)_{-\xi}\right\}^\dagger,
\end{equation}
\begin{equation}
\label{eq:ximappm2}
(O_F)_{-\xi}=\left\{(O_F^\dagger)_{\xi}\right\}^\dagger.
\end{equation}
\end{subequations}
The mapping becomes the Hermitian type when $\xi=0$ and in other cases the non-Hermitian type. 

There is a one-to-one correspondence between the fermion subspace projected by $\hat T_F$ and the boson subspace by $\hat T_B$.
For the state vectors, $\vert\psi\rangle$ and $\vert\psi'\rangle$, which belong to the fermion subspace projected by $\hat T_F$,
\begin{equation}
\label{eq:mteqh}
\begin{array}{lll}
\langle\psi\vert O_F\vert\psi'\rangle
&=&\langle\psi\vert\hat T_F O_F\hat T_F\vert\psi'\rangle
\\
&=&\langle\psi\vert\hat U_{-\xi}^\dagger U_\xi O_FU_{-\xi}^\dagger U\vert\psi'\rangle
\\
&=&{}_{-\xi}(\psi\vert(O_F)_\xi\vert\psi')_\xi,
\end{array}
\end{equation}
that is, the matrix elements of the fermion subspace become equal to those of the corresponding boson subspace.

The relation
\begin{equation}
{}_{\xi}(\psi\vert(O_F)_{-\xi}\vert\psi')_{-\xi}={}_{-\xi}(\psi\vert(O_F)_\xi\vert\psi')_\xi
\end{equation} 
holds, therefore it is sufficient to treat the case $\xi\ge 0$.

The mapping of the product of the fermion operators does not generally result in the product of the mapped fermion operators.
That is 
\begin{equation}
\label{eq:ximapzbar22}
(O_FO'_F)_{\xi}\neq (O_F)_{\xi}(O'_F)_{\xi},
\end{equation}
and therefore, the commutation relations of the fermion operators are mapped as
\begin{equation}
([O_F, O'_F])_\xi=(O_FO'_F)_\xi-(O'_FO_F)_\xi\neq[(O_F)_\xi, (O'_F)_\xi],
\end{equation}
while under the approximation $O_FO'_F\approx O_F\hat T_FO'_F$, 
\begin{equation}
\label{eq:ximapzbar22aprox}
(O_FO'_F)_{\xi}\approx (O_F)_\xi(O'_F)_\xi,
\end{equation}
and 
\begin{equation}
([O_F, O'_F])_\xi\approx [(O_F)_\xi, (O'_F)_\xi]
\end{equation}
hold.
The conventional practical boson expansion methods use this approximation.
If this approximation holds, it is sufficient to map the phonon and scattering operators, otherwise, it becomes necessary to obtain the mapping of the product of these fermion operators.

\subsection{Rewriting the mapping operator with the norm operator}
We utilize the norm operator \cite{KT83}, whose definition is
\begin{subequations}
\begin{equation}
\label{eq:normop}
\hat Z=\sum_{N=0}^{N_{max}}\hat Z(N),
\end{equation}
\begin{equation}
\label{eq:normop2}
\begin{array}{lll}
\hat{Z}(N)&=&\displaystyle\sum_{t t'}\vert N, t)\langle N;
t\vert N; t'\rangle ( N; t' \vert
\\
&=&
\displaystyle\sum_{t_1\leq\cdots \leq t_N}\sum_{t'_1\leq\cdots
\leq t'_N}\vert t_1 \cdots t_N)\langle t_1\cdots t_N\vert
t'_1\cdots t_N\rangle (t'_1\cdots t'_N \vert.
\end{array}
\end{equation}
\end{subequations}

The basis vectors of Eq. (\ref{eq:bbsistr}) become eigenvectors of $\hat Z(N)$ and $\hat Z$:
\begin{equation}
\label{eq:eigeneqnrop1}
\hat Z(N)\vert N; a))=z_a(N)\vert N; a)), \quad \hat Z\vert N; a))=z_a(N)\vert N; a)).
\end{equation}
We obtain the spectral decomposition of $\hat Z(N)$ as
\begin{equation}
\label{eq:nropspec}
\hat{Z}(N) = \sum_{a\neq a_0}\vert N; a))z_a(N)((N; a\vert.
\end{equation}
Using this, functions of $\hat Z(N)$ are defined by
\begin{equation}
f(\hat Z(N))=\sum_{a\neq a_0}\vert N; a))f(z_a(N))((N; a\vert,
\end{equation}
therefore
\begin{equation}
f(\hat Z)=\sum_{N=0}^{N_{max}}f(\hat Z(N)).
\end{equation}

Using the norm operator $\hat Z$, we express the mapping operator $U_\xi$ as
\begin{equation}
\label{eq:bmopzubar}
U_\xi=\hat Z^{\xi-\frac 12}\bar U,
\end{equation}
where $\bar U$ is a mapping operator whose definition is as follows:
\begin{subequations}
\label{eq:tildempop}
\begin{equation}
\label{eq:tildempop1}
\bar {U}=\sum_{N=0}^{N_{max}}\bar U(N),
\end{equation}
\begin{equation}
\label{eq:tildempop2}
\begin{array}{lll}
\bar U(N)&=&\displaystyle\sum_{t}\vert N; t)\langle N; t\vert
\\
&=&
\displaystyle\sum_{t_1\leq t_2\leq\cdots\leq t_N}\vert t_1 t_2\cdots
t_N)\langle t_1 t_2\cdots t_N\vert
\\
&=&\displaystyle\sum_{a\neq a_0}z_a^{\frac 12}\vert N; a))\langle\langle N; a\vert,
\end{array}
\end{equation}
\end{subequations}
which satisfies the following rerations,
\begin{subequations}
\label{eq:tildeurel}
\begin{equation}
\label{eq:tldeurel1}
\bar U\bar U^\dagger =\hat{Z},
\end{equation}
\begin{equation}
\label{eq:tildeurel2}
\bar U(N)\bar U(N)^\dagger =\hat{Z}(N).
\end{equation}
\end{subequations}

We denote the mapping by $\bar U$ as
\begin{subequations}
\label{eq:tildemap}
\begin{equation}
\label{eq:tildemap1}
\overline{\vert \psi)}= \bar U\vert\psi\rangle,\qquad\overline{(\psi\vert} =\langle\psi\vert \bar U^\dagger,
\end{equation}
\begin{equation}
\label{eq:tildemap2}
\overline{O_F}=\bar UO_F\bar U^\dagger.
\end{equation}
\end{subequations}
The mapping of Eqs. (\ref{eq:ximap}) is expressed as
\begin{subequations}
\label{eq:ximapzbar}
\begin{equation}
\label{eq:ximapzbar1}
\vert \psi')_{\xi}= \hat Z^{\xi-\frac 12}\overline{\vert\psi')},\qquad {}_{-\xi} (\psi\vert=\overline{(\psi\vert}\hat Z^{-\xi-\frac 12},
\end{equation}
\begin{equation}
\label{eq:ximapzbar2}
(O_F)_{\xi}=\hat Z^{\xi-\frac 12}\overline{O_F}\hat Z^{-\xi-\frac 12},
\end{equation}
\end{subequations}
which makes it clear that the different treatment of the norm operator produces different mapping.
The mapping of Eqs. (\ref{eq:ximap}) is also expressed as
\begin{subequations}
\label{eq:nhhr}
\begin{equation}
\label{eq:nhhr1}
\vert \psi')_{\xi}= \hat Z^{\xi}\vert\psi')_0,\qquad {}_{-\xi} (\psi\vert={}_0(\psi\vert\hat Z^{-\xi},
\end{equation}
\begin{equation}
\label{eq:nhhr2}
(O_F)_{\xi}=\hat Z^{\xi}(O_F)_0\hat Z^{-\xi},
\end{equation}
\end{subequations}
The mapping of the Hermitian type of $\xi=0$ and that of the non-Hermitian type of $\xi\neq 0$ transform one another by the similarity transformation operator composed of the power of the norm operator $\hat Z$.

\subsection{The case that the maximum phonon excitation number is 1}
In the case that the maximum phonon excitation number is 1, the results of the mapping are obtained concretely.
$\vert 0\rangle$ and $\vert\mu\rangle$ are an orthonormal system of the fermion space of even-quasiparticle excitations up to the two-quasiparticle excitations, which correspond to $\vert 0)$ and $\vert \mu)$, respectively, and it becomes possible to map all the excitation modes.
The mapping operator becomes as
\begin{equation}
\label{eq:ubarmax1}
U_\xi=\bar U=\vert 0)\langle 0\vert+\sum_\mu\vert\mu)\langle \mu\vert,
\end{equation}
and $\hat Z=\breve 1_B$.
As a result, the mapping operator has no dependence on $\xi$, and the mapping becomes Hermitian for any cases.
The projection operator onto the fermion subspace to be mapped is given by
\begin{subequations}
\label{eq:tftbmax1}
\begin{equation}
\label{eq:tfmax1}
\hat T_F=\vert 0\rangle\langle 0\vert +\sum_\mu\vert \mu\rangle\langle \mu\vert.
\end{equation}
The projection operator onto the physical subspace, which has a one-to-one correspondence to the fermion subspace, becomes as
\begin{equation}
\label{eq:tbmax1}
\hat T_B=\breve 1_B,
\end{equation}
\end{subequations}
which indicates that the ideal boson states are the physical state vectors, having one-to-one correspondences to the fermion state vectors.

The following relations:
\begin{subequations}
\label{eq:formulamax1}
\begin{equation}
\label{eq:formulaxtmax1}
X_\mu U_{-\xi}^\dagger =\vert 0\rangle(\mu\vert =U_{-\xi}^\dagger b_t\breve 1_B,
\end{equation}
\begin{equation}
\label{eq:formulaxtdaggermax1}
U_\xi X_\mu^\dagger =\vert \mu)\langle 0\vert =\breve 1_Bb_\mu^\dagger U_\xi,
\end{equation}
\begin{equation}
\label{eq:formulabqmax1}
B_qU_{-\xi}^\dagger =\sum_{\mu'}\sum_\mu\Gamma^{\mu'\mu}_q\vert\mu\rangle(\mu'\vert,
\end{equation}
\end{subequations}
hold, and we obtain
\begin{subequations}
\label{eq:bexpmax1}
\begin{equation}
\label{eq:bexpxtmax1}
(X_\mu)_\xi=\vert 0)(\mu\vert =\breve 1_B(X_\mu)_B\breve 1_B=(X_\mu)_B\breve 1_B,\quad (X_\mu)_B=b_\mu,
\end{equation}
\begin{equation}
\label{eq:bexpxtdaggermax1}
(X_\mu^\dagger)_\xi=\vert\mu)(0\vert =\breve 1_B(X_\mu^\dagger)_B\breve 1_B=\breve 1_B(X_\mu^\dagger)_B,\quad (X_\mu^\dagger)_B=b_\mu^\dagger,
\end{equation}
\begin{equation}
\label{eq:bexpbqmax1}
\begin{array}{c}
(B_q)_\xi=\displaystyle\sum_{\mu'}\sum_\mu\Gamma^{\mu'\mu}_q\vert\mu)(\mu'\vert=\breve 1_B(B_q)_B\breve 1_B=\breve 1_B(B_q)_B=(B_q)_B\breve 1_B,
\\
 (B_q)_B=\displaystyle\sum_{\mu \mu'}\Gamma^{\mu' \mu}_qb_\mu^\dagger b_{\mu'}.
\end{array}
\end{equation}
\end{subequations}

The product of the operators becomes as follows:
\begin{subequations}
\label{eq:bexpproductmax1}
\begin{equation}
(O_FX_\mu)_\xi=(O_F)_\xi(X_\mu)_\xi=\breve 1_B(O_F)_B\breve 1_B(X_\mu)_B\breve 1_B=\breve 1_B(O_F)_B(X_\mu)_B\breve 1_B,
\end{equation}
\begin{equation}
(X_\mu^\dagger O_F)_\xi=(X_\mu^\dagger)_\xi(O_F)_\xi=\breve 1_B (X_\mu^\dagger)_B\breve 1_B(O_F)_B\breve 1_B=\breve 1_B(X_\mu^\dagger)_B(O_F)_B\breve 1_B,
\end{equation}
\end{subequations}
therefore we can obtain the mapping of the product of $X_\mu^\dagger$,$X_\mu$, and $B_q$ by arranging them in normal order.

The commutation relations of $(X_\mu^\dagger )_B$, $(X_\mu)_B$, and $(B_q)_B$ become as follows:
\begin{subequations}
\label{eq:combapp}

\begin{equation}
\label{eq:combapp1}
[ (X_\mu)_B, (X_{\mu'}^\dagger)_B ]=\delta_{\mu,\mu'}
\end{equation}

\begin{equation}
\label{eq:combapp2}
[ (B_q)_B,( X_\mu^\dagger)_B]=\sum_{\mu'}\Gamma^{\mu\mu'}_q(X_{\mu'}^\dagger)_B.
\end{equation}

\begin{equation}
\label{eq:combapp3}
[ (X_\mu)_B, (B_q )_B]=\sum_{\mu'}\Gamma^{\mu'\mu}_q(X_{\mu'})_B,
\end{equation}
\end{subequations}
which are equal to the results of the boson approximation.

From the above, when the maximum number of phonons is $1$, by arranging the phonon creation and annihilation operators and the scattering operators in normal order and replacing them with $(X_\mu^\dagger)_B$, $(X_\mu)_B$, and  $(B_q)_B$, respectively, then the fermion subspace is completely mapped onto the boson subspace projected by $\breve 1_B$.

In this way, the boson approximation has been established as the boson mapping whose maximum phonon excitation number is 1.

\section{Boson expansions}
When the phonon excitation number exceeds 1, it becomes necessary to perform boson expansions. Hereafter, we express the order of magnitude of $\Gamma_q^{\mu\mu'}$ as $O(\Gamma)$.

\subsection{Boson expansions as small parameter expansions}
In the case that the phonon excitation number exceeds 1, sums of $\Gamma_q^{\mu\mu'}$ in terms of phonon excitation modes appear in the coefficient of the boson expansions. 
Whether
\color{blue} we can regard them as 
\color{black}
small or not depends on how to sum them up. As a result, in some cases, small parameter expansions become to fail, and the boson approximation does not become the zeroth-order approximation of the boson expansions.

\color{blue}
We investigate the two-phonon norm matrix composed of all the phonon excitation modes in detail and reveal the mechanism, which is essential for the boson expansion methods, of how we should construct the fermion subspace to be mapped from the whole fermion space.
\color{black}

First, we deal with the case where we take up all phonon excitation modes for constructing multi-phonon state vectors.
The multi-phonon norm matrix elements,
\begin{equation}
\label{eq:2mumu}
\langle\langle\langle\mu'_1\mu'_2\vert\mu_1\mu_2\rangle\rangle\rangle=(((\mu'_1\mu'_2\vert\mu_1\mu_2)))-Y(\mu'_1\mu_1\mu_2\mu'_2),
\end{equation}
consist of the corresponding boson norm matrix elements and $Y(\mu'_1\mu_1\mu_2\mu'_2)$, which express the effect of the Pauli exclusion principle.
We estimate the order of magnitude of $Y(\mu_1\mu_2\mu_3\mu_4)$ as $O(\Gamma^2)$, and the effect of the Pauli exclusion principle becomes small.
On the other hand, from Eq. (\ref{eq:complete}), which is the completeness condition of the Tamm-Dancoff phonon amplitudes, we derive the following \cite{KT72}
\begin{equation}
\label{eq:yymumusum}
\sum_{\mu\mu'}Y(\mu'_1\mu\mu'\mu'_2)Y(\mu_1\mu\mu'\mu_2)=4(((\mu'_1\mu'_2\vert\mu_1\mu_2)))-2Y(\mu'_1\mu_1\mu_2\mu'_2).
\end{equation}
The naive estimation with no consideration of the effect of sums of the left side of this equation is $ O (\Gamma^4) $, and the right side is $ O (1) $ or at most $ O (\Gamma^2) $, which becomes contradictory. We can not simply regard the left-hand side as small.
For investigation, we rewirte Eq. (\ref{eq:yymumusum}) as
\begin{equation}
\label{eq:Ymatrixrel}
\mathbf{Y}^2=2\mathbf{1}-\mathbf{Y},
\end{equation}
where $\mathbf{1}$ is a matrix whose elements are $(\mu'_1\mu'_2\vert\mu_1\mu_2)$ and $\mathbf{Y}$ is a matrix whose elements are $\mathcal{N}_B^{-\frac 12}(\mu'_1\mu'_2)Y(\mu'_1\mu_1\mu_2\mu'_2)\mathcal{N}_B^{-\frac 12}(\mu_1\mu_2)$.
The eigenvalues of $\mathbf{Y}$ are $-2$ and $1$. 
Denoting a 2-phonon norm matrix whose matrix elements are $\langle\mu'_1\mu'_2\vert\mu_1\mu_2\rangle$ as $\mathbf{Z_F}$, 
\begin{equation}
\mathbf{Z_F}=\mathbf{1}-\mathbf{Y},
\end{equation}
is obtained from Eq. (\ref{eq:2mumu}), and  $\mathbf{Z_F}$ has $0$ and $3$ as its eigenvalues, which quite deviate from $1$. Even if $\Gamma_q^{\mu\mu'}$ are small, the sum of them in terms of all phonon excitation modes does not become small. Therefore we should not treat all phonon excitation modes equally, and we must take care how to sum up small parameters $\Gamma_q^{\mu\mu'}$.

From the above consideration, it is necessary to restrict the excitation modes of the phonons constituting the multi-phonon state vectors for performing small parameter expansions.
We should treat
\begin{equation}
\label{eq:2tt}
\langle\langle\langle t'_1t'_2\vert t_1t_2\rangle\rangle\rangle=(((t'_1t'_2\vert t_1t_2)))-Y(t'_1t_1t_2t'_2)
\end{equation}
instead of Eq. (\ref{eq:2mumu}) and
\begin{equation}
\label{eq:yymumusumt}
\sum_{\mu\mu'}Y(t'_1\mu\mu't'_2)Y(t_1\mu\mu't_2)=4(((t'_1t'_2\vert t_1t_2)))-2Y(t'_1t_1t_2t'_2)
\end{equation}
instead of Eq(\ref{eq:yymumusum}).
We choose $\{t\} $ so that the $t$-sums do not affect the estimation. 
Applying this estimation to Eq. (\ref{eq:yymumusumt}), $\sum_{tt'}Y(t_1tt't_2)Y(t'_1tt't'_2)\sim O(\Gamma^4)$ holds, and we obtain
\begin{equation}
\label{eq:yy}
\sum_{\bar t\bar t'}Y(t_1\bar t\bar t't_2)Y(t'_1\bar t\bar t't'_2)+2\sum_{t\bar t}Y(t_1t\bar tt_2)Y(t'_1t\bar tt'_2)=4(((t'_1t'_2\vert t_1t_2)))-2Y(t'_1t_1t_2t'_2)+O(\Gamma^4).
\end{equation}
This left side includes the parts where the naive estimation fails. We should estimate the sum of $\bar t$ ($\bar t$-sum) another way.

To find out more about $\bar t$-sum, we take up $\sum_\mu Y(t_1t_2t_3\mu)\Gamma_q^{\mu t_4}$.  Because we choose $\{t\}$ so that $\sum_tY(t_1t_2t_3t)\Gamma_q^{tt_4}\sim O(\Gamma^3)$ hold, then we obtain
\begin{equation}
\sum_\mu Y(t_1t_2t_3\mu)\Gamma_q^{\mu t_4}=\sum_{\bar t}Y(t_1t_2t_3\bar t)\Gamma_q^{\bar tt_4}+O(\Gamma^3).
\end{equation}
While
\begin{equation}
\label{eq:musumyg}
\sum_\mu Y(t_1t_2t_3\mu)\Gamma_q^{\mu t_4}=\sum_{q'q''}\sum_{\alpha\beta\gamma}\varphi_q(\alpha\beta)\varphi_{q'}(\gamma\alpha)\varphi_{q''}(\gamma\alpha)(\Gamma_{q'}^{t_1t_2}\Gamma_{q''}^{t_3t_4}+\Gamma_{q'}^{t_1t_3}\Gamma_{q''}^{t_2t_4}),
\end{equation}
hold, which indicates that the order of the right-hand side is $O(\Gamma^2)$. Therefore the estimation of  $\bar t$-sum becomes as
\begin{equation}
\label{eq:musummygordr}
\sum_{\bar t} Y(t_1t_2t_3\bar t)\Gamma_q^{\bar tt_4}\sim O(\Gamma^2),
\end{equation}
which indicates that if we take a single $\bar t$-sum, we should estimate its order of magnitude by one order lower.
Based on this evaluation, we evaluate
\begin{equation}
\sum_{t\bar t}Y(t_1t\bar tt_2)Y(t'_1t\bar tt'_2)\sim O(\Gamma^3),
\end{equation}
therefore Eq. (\ref{eq:yy}) becomes
\begin{equation}
\label{eq:yybb}
\sum_{\bar t\bar t'}Y(t_1\bar t\bar t't_2)Y(t'_1\bar t\bar t't'_2)=4(((t'_1t'_2\vert t_1t_2)))-2Y(t'_1t_1t_2t'_2)+O(\Gamma^3),
\end{equation}
which means that we fail the small-parameter evaluation when the double $\bar t$-sums connect the coefficients.

In this way, we should limit the types of phonon excitation modes that compose the multi-phonon state vector and select these modes so that $t$-sums do not affect the estimation. Besides, we should estimate the order of the magnitude of the coefficients connected by a single $\bar t $-sum one lower. Then the failure of the small parameter expansions is limited to the coefficients having the double $\bar t$-sums. Therefore we should construct boson expansions having no coefficients with the double $\bar t$-sums.

Each coefficient to be summed should be small enough so that no $t$-sum modifies the order estimation of each coefficient. 
The modification by $\bar t$-sums that occurs, as a result, suggests that we should not neglect each coefficient to be summed just because of its smallness. Therefore we should adopt the collective modes, which superpose various quasiparticle excitations, and the minimum necessary non-collective modes as those of $\{t \}$, as conventional practical boson expansion methods do.
Eq.  (\ref{eq:yybb}) becomes the necessary condition for the proper selection of $\{t\}$.

The conventional practical boson expansion methods have adopted approximations that neglect the contribution of the unselected phonons.
These approximations bring the closed-algebra approximation, which closes the commutation relations among the selected phonon operators.

DBET adopts the phonon-truncation approximation, which truncates the unselected phonons as follows:
\begin{equation}
\label{eq:phtapprox}
a_\alpha^\dagger a_\beta^\dagger \approx \sum_{t}\psi_t(\alpha\beta)X_t^\dagger.
\end{equation}
Substituting Eq.(\ref{eq:phtapprox}) into Eq.(\ref{eq:phononopc}), we obtain
\begin{equation}
\label{btph0}
X_{\bar t}^\dagger\approx 0,
\end{equation}
which bring the following commutation relations:
\begin{subequations}
\label{eq:algebraphtapprox}
\begin{equation}
\label{eq:algebraphtapprox1}
[ X_t, X_{t'}^\dagger ]=\delta_{t,
t'}-\sum_q\Gamma^{tt'}_qB_q,
\end{equation}

\begin{equation}
\label{eq:algebraphtapprox2}
[ B_q, X_t^\dagger
]\approx\sum_{t'}\Gamma^{tt'}_qX_{t'}^\dagger,
\end{equation}

\begin{equation}
\label{eq:algebraphtapprox3}
[ X_t, B_q ]\approx\sum_{t'}\Gamma^{t't}_qX_{t'},
\end{equation}
\end{subequations}
and the commutation relations among the adopted phonon operators are closed:
\begin{equation}
\label{eq:doublecomphtapprox}
[ [X_{t_1}, X_{t_2}^\dagger], X_{t_3}^\dagger] \approx -\sum_{t'}Y(t_1, t_2, t_3, t')X_{t'}^\dagger
\end{equation}
\color{blue}
We call an approximation that closes the phonon double commutation relations between chosen modes the closed-algebra approximation.
The choice of modes for it is arbitrary.
Eq. (\ref{eq:doublecomphtapprox}) becomes the closed-algebra approximation for $\{t\}$.
\color{black}
In this way, we can derive the closed-algebra approximation from the phonon-truncation approximation.
\color{blue}
The phonon-truncation approximation, however, makes the submatrix composed of the modes not truncated have at least one zero eigenvalues.
\color{black}
Substituting Eq.(\ref{btph0}) into Eq. (\ref{eq:2mumu}), we obtain
\begin{equation}
\label{eqyphtapprox}
Y(\mu'_1\mu_1\mu_2\mu'_2)\approx (((\mu'_1\mu'_2\vert\mu_1\mu_2)))\quad (\mu'_1, \mu'_2, \mu_1, {\mathrm or} \mu_2\in\{\bar t\}).
\end{equation}
Applying these to Eq. (\ref{eq:yymumusumt}), we obtain
\begin{equation}
\label{eq:yymumusumttcaa}
\sum_{tt'}Y(t'_1tt't'_2)Y(t_1tt't_2)\approx 4(((t'_1t'_2\vert t_1t_2)))-2Y(t'_1t_1t_2t'_2).
\end{equation}
Therefore,
\begin{equation}
\label{eq:Ydmatrixrel}
\mathbf{Y'}^2\approx 2\mathbf{1'}-\mathbf{Y'},
\end{equation}
\begin{equation}
\mathbf{Z'_F}=\mathbf{1'}-\mathbf{Y'},
\end{equation}
hold, where $\mathbf{Z'_F}$, $\mathbf{1'}$, and $\mathbf{Y'}$ are the matrices whose elements are $\langle t'_1t'_2\vert t_1t_2\rangle$, $(t'_1t'_2\vert t_1t_2)$, and $\mathcal{N}_B^{-\frac 12}(t'_1t'_2)Y(t'_1t_1t_2t'_2)\mathcal{N}_B^{-\frac 12}(t_1t_2)$, respectively.
We can not apply the estimation for the small parameter expansions, $\mathbf{Y'}^2\sim O(\Gamma^4)$, to Eq. (\ref{eq:Ydmatrixrel}), and it derives wrong result $\mathbf{Z'_F}\approx -\mathbf{1'}$.
Eq. (\ref{eq:Ydmatrixrel}) is satisfied only when $\{t\}$ is taken so large that $t$-sums and  $\bar t$-sums play the inverse roles to those for the small parameter expansions. 
Eq. (\ref{eq:phtapprox}) suggests that this reversing is originally necessary for this approximation to hold. That is, we need so many sorts of the selected phonons to reproduce the fermion pair excitations.
In this case, $\mathbf{Y}$ becomes as follows:
\begin{equation}
\mathbf{Y}\approx\left (
\begin{array}{@{\,}cc@{\,}}
\mathbf{Y'}&\mathbf{0'}
\\
\mathbf{0'}^T&\mathbf{1''}
\end{array}
\right ),
\end{equation}
where $\mathbf{0'}$ and $\mathbf{1''}$ are a zero matrix and an identity matrix, respectively. $\mathbf{0'}^T$ is a transposed matrix of $\mathbf{0'}$.
This $\mathbf{Y}$ satisfies Eq. (\ref{eq:Ymatrixrel}), from which we also obtain Eq. (\ref{eq:Ydmatrixrel}).
The eigenvalues of $\mathbf{Z'_F}$ are $0$ and $3$, which is the same as $\mathbf{Z_F}$.
\color{blue}
We call this approximation CAA I.
\color{black}

NOLCEXP claims that we can make the parts where $\bar t$-sums are taken small enough to be neglected with no appearance of zero eigenvalues of the norm matrix of the multi-phonon state vector by taking \{t\} appropriately \cite{KT83, SK88}. This prescription also derives the closed-algebra approximation Eq.(\ref{eq:doublecomphtapprox}).
We express $\mathbf {Y} $,
\begin{equation}
\label{eq:y'wy77}
\mathbf{Y}=\left (
\begin{array}{@{\,}cc@{\,}}
\mathbf{Y'}&\mathbf{W}
\\
\mathbf{W}^T&\mathbf{Y''}
\end{array}
\right ),
\end{equation}
calculate its square, and neglect its components that contain $\bar t$-sums, which becomes as follows:
\begin{equation}
\mathbf{Y}^2=\left (
\begin{array}{@{\,}cc@{\,}}
\mathbf{Y'}^2+\mathbf{W}\mathbf{W}^T&\mathbf{Y'}\mathbf{W}+\mathbf{W}\mathbf{Y''}
\\
\mathbf{W}^T\mathbf{Y'}+\mathbf{Y''}\mathbf{W}^T&\mathbf{W}^T\mathbf{W}+\mathbf{Y''}^2
\end{array}
\right )
\approx
\left (
\begin{array}{@{\,}cc@{\,}}
\mathbf{Y'}^2&\mathbf{Y'}\mathbf{W}
\\
\mathbf{W}^T\mathbf{Y'}&\mathbf{W}^T\mathbf{W}
\end{array}
\right ).
\end{equation}
Substituting these into Eq. (\ref{eq:Ymatrixrel}), we obtain the following:
\begin{subequations}
\label{eq:tlarge}
\begin{equation}
\label{eq:tlarge1}
\mathbf{Y'}^2\approx 2\mathbf{1'}-\mathbf{Y'},
\end{equation}
\begin{equation}
\mathbf{Y'}\mathbf{W}\approx -\mathbf{W},
\end{equation}
\begin{equation}
\mathbf{W}^T\mathbf{W}\approx 2\mathbf{1''}-\mathbf{Y''}.
\end{equation}
\end{subequations}
We again obtain Eq. (\ref{eq:Ydmatrixrel}) and we can not avoid the appearance of the zero in the eigenvalues of $\mathbf{Z'_F}$.
These equations are not satisfied under the estimation for the small parameter expansions because the left-hand sides include only $t$-sums and are estimated as $O(\Gamma^4)$, while the right-side hand becomes $O(\Gamma^2)$ or $O(1)$. It is only possible when $\{t\}$ is taken so large that $t$-sums and  $\bar t$-sums replace their roles with one another, as is the same in the phonon-truncation approximation.
\color{blue}
We call this approximation CAA I\hspace{-.1em}I.
\color{black}

In this way, it has become clear that the methods of neglecting the unselected phonon excitation modes proposed in the conventional practical boson expansion methods \color{blue} do not hold without large $\{t\}$, which is  \color{black} on the contrary to the case the small parameter expansions hold.
The same is true for the closed-algebra approximation derived from these methods.

Next, we investigate how it happens to derive the closed-algebra approximation \color{blue} of Eq. (\ref{eq:doublecomphtapprox}) \color{black} by setting $Y(t_1t_2t_3\bar t)\approx 0$, \color{blue}which is equivalent to that $\vert t_1 t_2 \rangle$ and $\vert t \bar t \rangle$ are approximately orthogonal to each other.
\color{black} 
In this case, $\mathbf{Y}$ becomes as follows:
\begin{equation}
\label{eq:caay}
\mathbf{Y}\approx\left (
\begin{array}{@{\,}ccc@{\,}}
\mathbf{Y'}&\mathbf{0^{(1)}}&\mathbf{U}
\\
\mathbf{0^{(1)}}^T&\mathbf{Y'''}&\mathbf{V}
\\
\mathbf{U}^T&\mathbf{V}^T&\mathbf{Y''''}
\end{array}
\right ),
\end{equation}
and  we obtain
\begin{equation}
\mathbf{Y}^2\approx\left (
\begin{array}{@{\,}ccc@{\,}}
\mathbf{Y'}^2+\mathbf{U}\mathbf{U}^T&\mathbf{U}\mathbf{V}^T&\mathbf{Y'}\mathbf{U}+\mathbf{U}\mathbf{Y''''}
\\
\mathbf{V}\mathbf{U}^T&\mathbf{Y'''}^2+\mathbf{V}\mathbf{V}^T&\mathbf{Y'''}\mathbf{V}+\mathbf{V}\mathbf{Y''''}
\\
\mathbf{U}^T\mathbf{Y'}+\mathbf{Y''''}\mathbf{U}^T&\mathbf{V}^T\mathbf{Y'''}+\mathbf{Y''''}\mathbf{V}^T&\mathbf{U}^T\mathbf{U}+\mathbf{V}^T\mathbf{V}+\mathbf{Y''''}^2
\end{array}
\right ).
\end{equation}
Substituting these into Eq. (\ref{eq:Ymatrixrel}), we obtain
\begin{equation}
\mathbf{Y'}^2+\mathbf{U}\mathbf{U}^T\approx 2\mathbf{1'}-\mathbf{Y'},
\end{equation}
which brings
\begin{equation}
\label{eq:tttbtb}
\sum_{tt'}Y(t'_1tt't'_2)Y(t_1tt't_2)+\sum_{\bar t\bar t'}Y(t'_1\bar t\bar t't'_2)Y(t_1\bar t\bar t't_2)\approx 4(((t'_1t'_2\vert t_1t_2)))-2Y(t'_1t_1t_2t'_2).
\end{equation}
This is not contradictory to the condition for small parameter expansions Eq. (\ref{eq:yybb}), while, we also obtain the relation, $\mathbf{U}\mathbf{V}^T\approx \mathbf{0^{(1)}}$, that is
\begin{equation}
\sum_{\bar t\bar t'}Y(t'_1\bar t\bar t't'_2)Y(t_1\bar t\bar t'\bar t_2)\approx 0,
\end{equation}
which is not correct under the assumption of the small parameter expansions, and Eq. (\ref{eq:Ymatrixrel}) is not satisfied. 
\color{blue}
One possibility for this formula to hold might be that $\{t\}$ is taken so large that the double $\bar t$-sums become negligible. 
\color{black}
It is nothing but the enlargement of $\{t\}$, and we again derive Eq. (\ref{eq:Ydmatrixrel}) from Eq. (\ref{eq:tttbtb}). $\mathbf{Y'''}^2+\mathbf{V}\mathbf{V}^T$ is, \color{blue} however, \color{black} also regarded as zero matrix, and $\mathbf{Y}$ does not satisfy Eq. (\ref{eq:Ymatrixrel}). 
\color{blue}Equation (\ref{eq:Ymatrixrel}) is a major premise that must be satisfied in any case. 
It would be unnatural that $\vert t_1t_2\rangle$ and $\vert\bar t_1\bar t_2\rangle$ does not become orthogonal each other when $\vert t_1t_2\rangle$ and $\vert t\bar t\rangle$ is orthogonal each other.
Therefore we introduce another condition that $\vert t_1t_2\rangle$, and $\vert \mu\bar t\rangle$ become approximately orthogonal to each other. $\mathbf{U}$ becomes approximately a zero matrix, and $\mathbf{Y}$ becomes as follows:
\begin{equation}
\label{eq:caay2}
\mathbf{Y}\approx\left (
\begin{array}{@{\,}cc@{\,}}
\mathbf{Y'}&\mathbf{0'}
\\
\mathbf{0'}^T&\mathbf{Y''}
\end{array}
\right ),
\end{equation}
which is nothing but Eq. (\ref{eq:y'wy77}) with inputting $\mathbf{W}\approx\mathbf{0'}$.
Eq. (\ref{eq:Ymatrixrel}) becomes as
\begin{subequations}
\begin{equation}
\label{eq:Yd}
\mathbf{Y'}^2\approx 2\mathbf{1'}-\mathbf{Y'},
\end{equation}
\begin{equation}
\label{eq:Yddd}
\mathbf{Y''}^2\approx 2\mathbf{1''}-\mathbf{Y''},
\end{equation}
\end{subequations}
We obtain  Eq. (\ref{eq:Ydmatrixrel}) here again.
We call this approximation CAA I\hspace{-.1em}I\hspace{-.1em}I.

CAA I, I\hspace{-.1em}I, and  I\hspace{-.1em}I\hspace{-.1em}I have nesting structure:
Eq. (\ref{eq:doublecomphtapprox}) corresponds to Eq. (\ref{eq:doublecom}), and
Eq. (\ref{eq:Ydmatrixrel}), Eq. (\ref{eq:tlarge1}), and Eq. (\ref{eq:Yd}) correspond to Eq. (\ref{eq:Ymatrixrel}). 
By this nesting structure, the case of successful application of CAA I, I\hspace{-.1em}I, and I\hspace{-.1em}I\hspace{-.1em}I compels $\mathbf{Z'_F}$ to have zero eigenvalues and makes small parameter expansions impossible.
For the success of CAA I and I\hspace{-.1em}I, $\{t\}$ should include so many sorts of phonon excitation modes as if $\{t\}$ were regarded as $\{\mu\}$.
Therefore, the application of CAA I and I\hspace{-.1em}I fails when $\{t\} $ consists of limited excitation modes, that is, collective and not so many non-collective excitation modes.
CAA I\hspace{-.1em}I\hspace{-.1em}I also does not hold under the usual selection of $\{t\}$.
Although the state vectors consisting of different quasi-particle pairs are orthogonal to each other, those consisting of two phonons are not generally because the phonons consist of superpositions of quasi-particle pairs.  $\{t\}$ includes the collective excitation modes, and the phonon with these modes especially consists of more sorts of quasi-particle pair operators than those with the non-collective modes, therefore CAA I\hspace{-.1em}I\hspace{-.1em}I does not hold.

There remains the possibility that we can apply CAA I\hspace{-.1em}I\hspace{-.1em}I in the case of adopting more kinds of phonon excitation modes from $\{\bar t\}$ in addition to $\{t\}$.
That is, constructing $\{\tau\}$ by $\{t\}$ and a part of $\{\bar t\}$, and $\{\bar\tau\}$ by remaining modes, we have the possibility that make $\vert\tau\tau'\rangle$ become approximately orthogonal to $\vert\mu\bar\tau\rangle$.
If it holds, we can omit the excitation modes belonging to $\{\bar\tau\} $ from $\bar t$-sums.  We should, however, due to the nesting structure, make $\{\tau\}$ large enough compared to $\{t\}$ for the success of the small parameter expansions.
Considering the difference in properties between the collective excitation modes and the non-collective excitation modes,  we could find such $\{\tau\}$ that CAA I\hspace{-.1em}I\hspace{-.1em}I can be applied for, where the higher collectivity the excitation modes in $\{t\}$ acquire, the larger $\{\tau\}$ must become.

Although it seems to be sufficient only to assume that $\vert tt'\rangle$ and $\vert \mu\bar\tau\rangle$ are approximately orthogonal to each other  for removing those belonging to $\{\bar\tau\}$ from $\bar t$-sums,  Eq. (\ref{eq:yymumusum}) requires, in this case, the following equations are satisfied:
\begin{equation}
\label{eq:w1y3}
\sum_{\tau\tau'}Y(t\tau\tau't')Y(\mu\tau\tau'\bar\tau)\approx 0.
\end{equation}
It would be a natural assumption that $\vert\tau\tau'\rangle$ is not generally orthogonal to each other because $\{t\}$ in $\{\tau\}$ includes the collective excitation modes, which should overlap others belonging to $\{\tau\}$ and become the leading cause of generating $\bar t$-sum so that the small parameter expansions hold.
 Therefore $Y(t\tau\tau't')$ should not generally become zero, and Eq. (\ref{eq:w1y3}) requires that $Y(\mu\tau\tau'\bar\tau)$ should become approximately zero.
As a result, $\vert\tau\tau'\rangle$ should also become approximately orthogonal to $\vert \mu\bar\tau\rangle$.
It is nothing but the realization of the successful application of CAA I\hspace{-.1em}I\hspace{-.1em}I to $\{\tau\}$.
Its realization requires that not only the collective modes belonging to $\{t\}$ but also the non-collective modes belonging to $\{\tau\}$ have approximately no overlap with those belonging to $\{\bar\tau\}$, which is a natural and possible requirement.
The property of the phonon excitation modes makes the applicability of  CAA I\hspace{-.1em}I\hspace{-.1em}I depend mainly on the collectivity of the collective modes.
The higher the collectivity of the modes in $\{t\}$ becomes, the larger $\{\tau\}$ CAA I\hspace{-.1em}I\hspace{-.1em}I would need.
As is the case of  CAA I\hspace{-.1em}I\hspace{-.1em}I, CAA I and I\hspace{-.1em}I also should not be applied to $\{t\}$ but be applied to $\{\tau\}$ which includes enough kinds of the phonon excitation modes not adopted as boson excitation modes.
Although the closed-algebra approximation can not remove $\bar t$-sums themselves properly, it is possible to remove the modes not effective from $\bar t$-sums adequately.

In the first place, the boson expansion method is related to the kinematical problem handling the effect of the Pauli exclusion principle.
The above investigation of the closed-algebra approximation using a two-phonon norm matrix reveals the essence of the boson expansion method.
The collective phonon excitation modes overlap many sorts of other phonon excitation modes, while the non-collective ones do not so many.
The closed-algebra approximation applied for only a few phonon excitation modes including the collective ones fails due to the collective ones.
The success of the closed-algebra approximation needs not only the phonon excitation modes selected for constructing the multi-phonon state vectors including the collective ones but also the non-collective ones that are not treated as the boson excitations themselves but contribute to the coefficients of the boson expansions.
The small parameter expansions need that only a few kinds of modes should be selected as the phonon excitation modes having correspondence to the boson excitations and some of them should be the collective modes that have high enough collectivity.
The higher collectivity the collective excitation modes acquire, the more sorts of modes having no correspondence to bosons should be introduced to treat the effect of the Pauli exclusion principle accurately, which makes the small parameter expansions better.

From now on, we denote CAA-C as the closed-algebra approximation applying to $\{t\}$, which corresponds to the case that the conventional boson expansion methods do.
\color{black}

At this point, we again check the estimation for the small parameter expansions.
Substituting Eq. (\ref{eq:y'wy77}) into Eq. (\ref{eq:Ymatrixrel}), we obtain following relations:
\begin{subequations}
\label{eq:tsmall}
\begin{equation}
\mathbf{Y'}^2+\mathbf{W}\mathbf{W}^T=2\mathbf{1'}-\mathbf{Y'},
\end{equation}
\begin{equation}
\mathbf{Y'}\mathbf{W}+\mathbf{W}\mathbf{Y''}=-\mathbf{W},
\end{equation}
\begin{equation}
\mathbf{W}^T\mathbf{W}+\mathbf{Y''}^2=2\mathbf{1''}-\mathbf{Y''}.
\end{equation}
\end{subequations}
Apply the estimation to the terms that do not contain $\bar t$ -sum and obtain:
\begin{subequations}
\label{eq:y'y''wspe}
\begin{equation}
\label{eq:y'y''wspe1}
\mathbf{W}\mathbf{W}^T=2\mathbf{1'}-\mathbf{Y'}+O(\Gamma^4),
\end{equation}
\begin{equation}
\label{eq:y'y''wspe2}
\mathbf{W}\mathbf{Y''}=-\mathbf{W}+O(\Gamma^4),
\end{equation}
\begin{equation}
\label{eq:y'y''wspe3}
\mathbf{Y''}^2=2\mathbf{1''}-\mathbf{Y''}+O(\Gamma^4).
\end{equation}
\end{subequations}
The left-hand sides of these equations contain the single and the double $\bar t$-sums and the latter makes the left-hand sides as large as the right-hand sides, which makes these equations hold.
We can derive Eq. (\ref{eq:yybb}) from Eq. (\ref{eq:y'y''wspe1}). Eq. (\ref{eq:y'y''wspe3}) indicates that the matrix $\mathbf{Z''_F}=\mathbf{1''}-\mathbf{Y''}$ has zero eigenvalues.
It is interesting to compare Eq. (\ref{eq:y'y''wspe}) with Eq. (\ref{eq:tlarge}), where $t$-sums and $\bar t$-sums replace their roles one another.
Representing the 2-phonon norm matrix with the symbols used so far, $\mathbf{Z_F}$ is expressed as
\begin{equation}
\mathbf{Z_F}=\left (
\begin{array}{@{\,}cc@{\,}}
\mathbf{Z'_F}&-\mathbf{W}
\\
-\mathbf{W}^T&\mathbf{Z''_F}.
\end{array}
\right ),
\end{equation}
$\mathbf{Z_F}$ has necessarily zero eigenvalues \color{blue} no matter how we set $\psi_\mu(\alpha\beta)$.
\color{black}
For the success of the small parameter expansion, we should not take up and treat $\mathbf {Z_F} $ itself, but take up the submatrix $\mathbf {Z'_F} $, and make it have no zero eigenvalues under the condition that Eq. (\ref{eq:Ymatrixrel}) holds.
\color{blue}
Its success depends on how to treat $\psi_\mu(\alpha\beta)$.
We should presuppose Eq. (\ref{eq:Ymatrixrel}) for the order estimation for the small parameter expansions. 
\color{black}
Setting the submatrix $\mathbf {Z'_F} $ having no zero eigenvalues results in the submatrix $\mathbf {Z''_F} $ have zero eigenvalues. The proper
\color{blue}
$\psi_t(\alpha\beta)$
\color{black} makes the small parameter expansions hold. \color{blue}
The convergence problem occurs by a lack of collectivity of the collective modes and the inclusion of too many sorts of the non-collective mode into $\{t\}$. 
In the case where the small parameter expansions hold, CAA-C results in neglecting the effect of the Pauli exclusion principle improperly by neglecting the $\bar t$-sums.
\color{black}

\subsection{On the physical state vectors}
It is essential for boson expansion methods to be practical that the ideal boson state vectors, which reflect no effect of the Pauli exclusion principle, become physical. How to select the phonon excitation modes and set the upper limit of the number of excitations determines not only whether the small-parameter expansions are possible or not but also whether the ideal boson state vectors are physical or not.

$\hat T_B$ of Eq. (\ref{eq:proptb})  is the projection operator onto the physical boson subspace, and $\vert N; a))$ ($a\neq a_0$) of Eq. (\ref{eq:bbsistr}) are orthonormal basis vectors which span the physical boson subspace. In case the norm operator  $\hat Z$ has zero eigenvalues,
\begin{equation}
\hat T_B\vert N, t)=\sum_{N=0}^{N_{max}}\sum_{a\neq 0}\vert N;a))((N; a\vert N; t)\neq\vert N; t),
\end{equation}
hold, therefore the ideal boson state vectors $\vert N; t)$ do not become physical. The physical state vectors are $\vert N; a))\quad (a\neq a_0)$, whose structures are complicated by reflecting the Pauli exclusion principle.
The proper selection of the phonon excitation modes $\{t\}$ and the maximum phonon excitation number $N_{max}$, however, enables the norm operator $\hat Z$ to have no zero eigenvalues, and
\begin{equation}
\label{eq:tbN eq1bN}
\hat T_B(N)=\hat1_B(N)\quad (N\leq N_{max}),
\end{equation}
hold from Eq. (\ref{eq:eigenvcon2}). Therefore
\begin{equation}
\label{eq:tb eq 1b}
\hat T_B=\breve 1_B
\end{equation}
holds, and the ideal boson state vectors, which do not bear the effect of the Pauli exclusion principle, become physical.

In the case
\color{blue}
where CAA-C is 
\color{black}
successfully applied, the eigenvalues of the multi-phonon-norm matrix include those regarded as 0 at the stage of 2-phonon excitation, and we, then, can not use the ideal boson state vectors as physical. Conversely, in the case that the ideal boson state vectors are physical, we can not apply these approximations successfully, which results in neglecting the important terms of the expansions. The same is true for the closed-algebra approximation derived from these methods.

\subsection{Formulas for the boson expansions}
In the case of $N_ {max} \ge 2$, we can obtain  the boson expansions of the mapped fermion operators using 
\begin{equation}
\begin{array}{lll}
\label{eq:overlineunexp}
\displaystyle\bar U(N)&=&\displaystyle\sum_{t_1\leq t_2\leq\cdots\leq t_N}\vert t_1 t_2\cdots t_N)\langle t_1t_2\cdots t_N\vert
\\
&=&\displaystyle\sum_{t_1t_2\cdots t_N}\frac{{\mathcal N}_B(t_1t_2\cdots t_N)}{N!}\vert t_1 t_2\cdots t_N)\langle t_1t_2\cdots t_N\vert
\\
&=&\displaystyle\sum_{t_1t_2\cdots t_N}\frac{1}{N!}\vert t_1 t_2\cdots t_N)))\langle\langle \langle t_1t_2\cdots t_N\vert,
\end{array}
\end{equation}
which gives the following series of formulas:
\begin{subequations}
\label{eq:ubarop}
\begin{equation}
\bar U(N)X_{t'}=(X_{t'})_D\bar U(N+1)\quad (N\geq 0),
\end{equation}
\begin{equation}
\begin{array}{lll}
\bar U(1)X_t^\dagger&=&(X_t^\dagger)_D\bar U(0),
\\
\bar U(N+1)X_t^\dagger&=&(X_t^\dagger)_D\bar U(N)
\\
&&-\displaystyle\frac12\sum_{t_1t_2}\sum_{\bar t_1'}Y(tt_1t_2\bar t'_1)b_{t_1}^\dagger b_{t_2}^\dagger\bar U(N-1)X_{\bar t'_1}\quad (N\geq 1),
\end{array}
\end{equation}

\begin{equation}
\begin{array}{lll}
\bar U(0)B_q&=&0,
\\
\bar U(N)B_q&=&(B_q)_D\bar U(N)+\displaystyle\sum_t\sum_{\bar t'}\Gamma_q^{\bar t't}b_t^\dagger\bar U(N-1)X_{\bar t'}\quad (N\geq 1),
\end{array}
\end{equation}
\end{subequations}

\begin{equation}
\begin{array}{lll}
\bar U(1)X_{\bar t}^\dagger&=&0,
\\
\bar U(N+1)X_{\bar t}^\dagger&=&\displaystyle -\frac12\sum_{t_1t_2}\sum_{t'_1}Y(\bar t t_1t_2t'_1)b_{t_1}^\dagger b_{t_2}^\dagger b_{t'_1}\bar U(N)
\\
&&\displaystyle -\frac 12 \sum_{t_1t_2}\sum_{\bar t'_1}Y(\bar t t_1t_2\bar t'_1)b_{t_1}^\dagger b_{t_2}^\dagger \bar U(N-1)X_{\bar t'_1}\quad (N\geq 1),
\end{array}
\end{equation}

where
\begin{subequations}
\label{eq:dbexp}
\begin{equation}
(X_{t'})_D=b_{t'},
\end{equation}
\begin{equation}
\label{eq:xdaggerd}
(X_t^\dagger)_D=b_t^\dagger-\frac 12\sum_{t_1t_2}\sum_{t'_1}Y(tt_1t_2t'_1)b_{t_1}^\dagger b_{t_2}^\dagger b_{t'_1},
\end{equation}
\begin{equation}
\label{eq:bqd}
(B_q)_D=\sum_t\sum_{t'}\Gamma_q^{t't}b_t^\dagger b_{t'}.
\end{equation}
\end{subequations}
Eqs. (\ref{eq:dbexp}) are the same as the boson expansions derived by DBET.  $(B_{\bar q})_D^\dagger =(B_q)_D$ holds.

From these formulas, we obtain
\begin{subequations}
\label{eq:uoun}

\begin{equation}
\overline{X_{t'}}(N) =(X_{t'})_D\hat Z(N+1)\qquad (N\geq 0),
\end{equation}

\begin{equation}
\label{eq:barxtdagn}
\begin{array}{lll}
\overline{X_t^\dagger}(0)&=&(X_t^\dagger )_D\hat Z(0)
\\
\overline{X_t^\dagger}(N)&=&(X_t^\dagger )_D\hat Z(N)
 -\displaystyle\frac 12\sum_{t_1t_2}\sum_{\bar t'_1}Y(tt_1t_2\bar t'_1)b_{t_1}^\dagger b_{t_2}^\dagger \overline{X_{\bar t'_1}}(N-1)\qquad (N\ge 1),
\end{array}
\end{equation}

\begin{equation}
\begin{array}{lll}
\label{eq:barbqn}
\overline{B_q}(0)&=&0,
\\
\overline{B_q}(N) &=&(B_q)_D\hat Z(N)+\displaystyle\sum_t\sum_{\bar t'}\Gamma_q^{\bar t't}b_t^\dagger\overline{X_{\bar t'}}(N-1)\qquad (N\ge 1),
\end{array}
\end{equation}
\end{subequations}

\begin{equation}
\label{eq:barxbtdagn}
\begin{array}{lll}
\overline{X_{\bar t}^\dagger}(0)&=&0,
\\
\overline{X_{\bar t}^\dagger}(N)
&=&-\displaystyle\frac 12\sum_{t_1t_2}\sum_{t'_1}Y(\bar tt_1t_2t'_1)b_{t_1}^\dagger b_{t_2}^\dagger b_{t'_1}\hat Z(N)
\\
&-&\displaystyle\frac 12\sum_{t_1t_2}\sum_{\bar t'_1}Y(\bar tt_1t_2\bar t'_1)b_{t_1}^\dagger b_{t_2}^\dagger \overline{X_{\bar t'_1}}(N-1)\qquad (N\ge 1),
\end{array}
\end{equation}
where we use the following diffinitions: $\overline{X_\mu}(N)=\overline U(N)X_\mu\overline U(N+1)^\dagger$, $\overline{B_q}(N)=\overline U(N)B_q\overline U(N)^\dagger$. $\overline{X_\mu^\dagger}(N)=\left (\overline{X_\mu}(N)\right )^\dagger$ and $\overline{B_q^\dagger}(N)=\left ({\overline B_q}(N)\right )^\dagger$ hold.

We obtain the boson expansion of $\hat Z(N)$ by using 
\begin{equation}
\label{eq:zn}
\hat Z(N)=\frac 1N\sum_t(X_t^\dagger)_D\hat Z(N-1)b_t-\frac 1{2N}\sum_{t_1t_2}\sum_{t'_1}\sum_{\bar t'}Y(t'_1t_1t_2\bar t')b_{t_1}^\dagger b_{t_2}^\dagger\overline{X_{\bar t'}}(N-2) b_{t'_1}\quad (N\geq 1),
\end{equation}
which is derived by applying
\begin{equation}
\bar U^\dagger (N) =\frac 1N\sum_tX_t^\dagger\bar U^\dagger (N-1)b_t\qquad (N\ge 1),
\end{equation}
obtained from Eq. (\ref{eq:overlineunexp}), to Eq. (\ref{eq:tldeurel1}), expressing $\hat Z(N)$ as
\begin{equation}
\label{eq:znf}
\hat Z(N)=\frac 1N\sum_t\overline{X_t^\dagger}(N-1)b_t\quad (N\geq 1),
\end{equation}
and substituting Eqs. (\ref{eq:barxtdagn}) into this.

We can obtain the boson expansions of $\overline{X_{\bar t'}}(N)$, $\overline{X_{\bar t}}^\dagger(N)$, and $\hat Z(N)$ from Eq. (\ref{eq:barxbtdagn}), their Hermitian conjugate, and Eq. (\ref{eq:zn}). Substituting these results into Eqs. (\ref{eq:uoun}) gives the boson expansions of $\overline{X_{t'}}(N)$, $\overline{X_t}^\dagger (N)$, and $\overline{B_q}(N)$.
There are no terms including the double $\bar t$-sums, then we can achieve the small parameter expansions.

\subsection{Small parameter expansions of the mapped fermion operators}
Here we perform the boson expansions of the mapped fermion operators.

Eq. (\ref{eq:ximapzbar2}) indicates that we can derive the boson expansions of $(O_F)_\xi$ from those of the norm operator $\hat Z$ and $\overline{O_F}$. We give the terms of the boson expansions up to the order of magnitude $O(\Gamma^4)$.

From Eq. (\ref{eq:barxbtdagn}), its Hermitian conjugate, and Eq. (\ref{eq:zn}), we find the recurrence formulas for obtaining the boson expansions of $\hat Z(N)$, $\overline{X_{\bar t'}}(N)$, and $\overline{X_{\bar t}^\dagger}(N)$ up to the desired order of magnitude. Each coefficient in the expansions has less than two $\bar t$-sums,  which guarantees convergence of expansions as small parameter expansions.
The recurrence formulas of $\hat Z(N)$ are as follows:
\begin{subequations}
\label{eq:znrec}
\begin{equation}
\label{eq:znrec01234}
\hat Z(N)=\sum_{k=1}^4\hat Z^{(k)}(N)+O(\Gamma^5);\quad \hat Z^{(k)}(N)\sim O(\Gamma^k),
\end{equation}
\begin{equation}
\label{eq:znrec0}
\hat Z^{(0)}(N)=\frac 1N\sum_tb_t^\dagger\hat Z^{(0)}(N-1)b_t,
\end{equation}
\begin{equation}
\label{eq:znrec1}
\hat Z^{(1)}(N)=0,
\end{equation}

\begin{equation}
\label{eq:znrec2}
\begin{array}{lll}
\hat Z^{(2)}(N)&=&\displaystyle\frac 1N\sum_tb_t^\dagger\hat Z^{(2)}(N-1)b_t
\\
&&-\displaystyle\frac 1{2N}\sum_{t_1t_2}\sum_{t'_1t'_2}Y(t'_2t_1t_2t'_1)b_{t_1}^\dagger b_{t_2}^\dagger b_{t'_1}\hat Z^{(0)}(N-1)b_{t'_2},
\end{array}
\end{equation}
\begin{equation}
\label{eq:znrec3}
\begin{array}{lll}
\hat Z^{(3)}(N)&=&\displaystyle\frac 1N\sum_tb_t^\dagger\hat Z^{(3)}(N-1)b_t
\\
&&+\displaystyle\frac 1{4N}\sum_{t_1t_2t_3}\sum_{t'_1t'_2t'_3}\sum_{\bar t}Y(t'_3t_1t_2\bar t)Y(\bar tt'_1t'_2t_3)b_{t_1}^\dagger b_{t_2}^\dagger\hat Z^{(0)}(N-1)b_{t_3}^\dagger b_{t'_1}b_{t'_2}b_{t'_3},
\end{array}
\end{equation}

\begin{equation}
\label{eq:znrec4}
\begin{array}{c}
\hat Z^{(4)}(N)=\displaystyle\frac 1N\sum_tb_t^\dagger\hat Z^{(4)}(N-1)b_t
-\displaystyle\frac 1{2N}\sum_{t_1t_2}\sum_{t'_1t'_2}Y(t'_2t_1t_2t'_1)b_{t_1}^\dagger b_{t_2}^\dagger b_{t'_1}\hat Z^{(2)}(N-1)b_{t'_2}
\\
-\displaystyle\frac 1{8N}\sum_{t_1t_2t_3t_4}\sum_{t'_1t'_2t'_3t'_4}\sum_{\bar t\bar t'}Y(t'_4t_1t_2\bar t)Y(\bar tt'_2t'_3\bar t')Y(\bar t't_3t_4t'_1)b_{t_1}^\dagger b_{t_2}^\dagger b_{t_3}^\dagger b_{t_4}^\dagger b_{t'_1}\hat Z^{(0)}(N-1)b_{t'_2}b_{t'_3}b_{t'_4}.
\end{array}
\end{equation}
\end{subequations}

The solution of Eq,(\ref{eq:znrec0}) is
\begin{equation}
\label{eq:znrec0sol}
\hat Z^{(0)}(N)=\frac 1{N!}\sum_{t_1t_2\cdots t_N}b_{t_1}^\dagger b_{t_2}^\dagger\cdots b_{t_N}^\dagger\hat Z(0)b_{t_1}b_{t_2}\cdots b_{t_N}=\hat 1_B(N).
\end{equation}
Substituting it into Eq. (\ref{eq:znrec2}) and using
\begin{equation}
\label{eq:b1bn}
b_t\hat1_B(N)=\hat 1_B(N-1)b_t,
\end{equation}
we obtain
\begin{subequations}
\begin{equation}
\label{eq:znrec2'}
\begin{array}{lll}
\hat Z^{(2)}(N)&=&\displaystyle\frac 1N\sum_tb_t^\dagger\hat Z^{(2)}(N-1)b_t
\\
&&-\displaystyle\frac 1{2N}\sum_{t_1t_2}\sum_{t'_1t'_2}Y(t'_2t_1t_2t'_1)b_{t_1}^\dagger b_{t_2}^\dagger\hat 1_B(N-2) b_{t'_1}b_{t'_2}.
\end{array}
\end{equation}
For finding the solution, assuming it as
\begin{equation}
\hat Z^{(2)}(N)=y^{(2)}(N)\sum_{t_1t_2}\sum_{t'_1t'_2}Y(t'_2t_1t_2t'_1)b_{t_1}^\dagger b_{t_2}^\dagger\hat 1_B(N-2) b_{t'_1}b_{t'_2},
\end{equation}
substituting this into the recurrent formula, and using
\begin{equation}
\sum_tb_t^\dagger\hat 1_B(N)b_t=(N+1)\hat 1_B(N+1),
\end{equation}we find
\begin{equation}
y^{(2)}(N)=\frac{N-2}Ny^{(2)}(N-1)-\frac1{2N}.
\end{equation}
This solution is found as $y^{(2)}(N)=-1/4$, and we obtain
\begin{equation}
\hat Z^{(2)}(N)=-\frac 14\sum_{t_1t_2}\sum_{t'_1t'_2}Y(t'_2t_1t_2t'_1)b_{t_1}^\dagger b_{t_2}^\dagger\hat 1_B(N-2) b_{t'_1}b_{t'_2}.
\end{equation}
\end{subequations}
Following the same procedure in order, we can obtain the solution of the recurrence formula for each order of magnitude. Organizing the solutions obtained in this way using Eq. (\ref{eq:b1bn}), we finally obtain
\begin{subequations}
\label{eq:znexp}
\begin{equation}
\label{eq:znor4}
\hat Z(N)=\mathcal{\hat Z}\hat 1_B(N)=\hat 1_B(N)\mathcal{\hat Z};\quad\mathcal{\hat Z}=\mathcal{\hat Z}^{(0)}+\mathcal{\hat Z}^{(2)}+\mathcal{\hat Z}^{(3)}+\mathcal{\hat Z}^{(4)}+O(\Gamma^5),
\end{equation}
\begin{equation}
\label{eq:zn0}
\mathcal{\hat Z}^{(0)}=\hat 1_B,
\end{equation}
\begin{equation}
\label{eq:zn2}
\mathcal{\hat Z}^{(2)}=-\frac 14\sum_{t_1t_2}\sum_{t'_1t'_2}Y(t'_2t_1t_2t'_1)b_{t_1}^\dagger b_{t_2}^\dagger b_{t'_1}b_{t'_2},
\end{equation}

\begin{equation}
\label{eq:zn3}
\mathcal{\hat Z}^{(3)}=\mathcal{\hat Z}^{(3)}_{out}=\frac 1{12}\sum_{t_1t_2t_3}\sum_{t'_1t'_2t'_3}\sum_{\bar t}Y(t'_3t_1t_2\bar t)Y(\bar tt'_1t'_2t_3)b_{t_1}^\dagger b_{t_2}^\dagger b_{t_3}^\dagger b_{t'_1}b_{t'_2}b_{t'_3},
\end{equation}

\begin{equation}
\label{eq:zn4}
\begin{array}{lll}
\mathcal{\hat Z}^{(4)}&=&\mathcal{\hat Z}_{in}^{(4)}+\mathcal{\hat Z}_{out}^{(4)}
\\
&\mathcal{\hat Z}_{in}^{(4)}&=\displaystyle\frac 1{12}\sum_{t_1t_2t_3}\sum_{t'_1t'_2t'_3}\sum_{t}Y(t'_1t_1t_2 t)Y(tt'_2t'_3t_3)b_{t_1}^\dagger b_{t_2}^\dagger b_{t_3}^\dagger b_{t'_1}b_{t'_2}b_{t'_3}
\\
&&+\displaystyle\frac 1{32}\sum_{t_1t_2t_3t_4}\sum_{t'_1t'_2t'_3t'_4}Y(t'_2t_1t_2t'_1)Y(t'_4t_3t_4t'_3)b_{t_1}^\dagger b_{t_2}^\dagger b_{t_3}^\dagger b_{t_4}^\dagger b_{t'_1}b_{t'_2}b_{t'_3}b_{t'_4},
\\
&\mathcal{\hat Z}_{out}^{(4)}&=-\displaystyle\frac 1{32}\sum_{t_1t_2t_3t_4}\sum_{t'_1t'_2t'_3t'_4}\sum_{\bar t\bar t'}Y(t'_4t_1t_2\bar t)Y(\bar tt'_2t'_3\bar t')Y(\bar t't_3t_4t'_1)b_{t_1}^\dagger b_{t_2}^\dagger b_{t_3}^\dagger b_{t_4}^\dagger b_{t'_1}b_{t'_2}b_{t'_3}b_{t'_4}.
\end{array}
\end{equation}
\end{subequations}
The ``out'' subscript denotes the term that CAA-C discards.
From these results, we can easily find the norm operator as
\begin{equation}
\label{eq:normopres}
\begin{array}{lll}
\hat Z&=&\mathcal{\hat Z}\breve 1_B=\breve 1_B\mathcal{\hat Z}=\breve 1_B\mathcal{\hat Z}\breve 1_B,
\\
&\mathcal{\hat Z}&=\hat 1_B+\mathcal{\hat Y},
\\
&&\mathcal{\hat Y}=\mathcal{\hat Y}_{in}+\mathcal{\hat Y}_{out},
\\
&&\quad\mathcal{\hat Y}_{in}=\mathcal{\hat Z}^{(2)}+\mathcal{\hat Z}_{in}^{(4)}+O(\Gamma^5),
\\
&&\quad\mathcal{\hat Y}_{out}=\mathcal{\hat Z}^{(3)}+\mathcal{\hat Z}_{out}^{(4)}+O(\Gamma^5).
\end{array}
\end{equation}
The $\xi$-th power of $\hat Z$ becomes
\begin{equation}
\label{eq:normopresp}
\begin{array}{l}
\hat Z^\xi=\mathcal{\hat Z}^\xi\breve 1_B=\breve 1_B\mathcal{\hat Z}^\xi,
\\
\quad\mathcal{\hat Z}^\xi=\hat 1_B+\xi\mathcal{\hat Y}+\displaystyle\frac 12\xi(\xi-1)\mathcal{\hat Y}^2+O(\Gamma^6).
\end{array}
\end{equation}

In this way, we have been able to find the boson expansion for the norm operator. 
\color{blue}This enables us to solve the eigenvalue problem of the norm operator Eq. (\ref{eq:eigeneqnrop1}) in the boson subspace used for solving the eigenvalue problem for the Hamiltonian.
Since the norm operator saves the number of bosons, we can solve its eigenvalue problem for each number of bosons excitations. In particular, it is important to confirm that the zero eigenvalues do not appear in the result solved for the maximum value of the phonon excitation number $ N_ {max} $, that is, the maximum value of the boson excitation number, which is much easier than to solve the eigenvalue problem of the Hamiltonian.
The boson expansion of the norm operator not only facilitates the boson expansions, as will be seen later but also makes it possible to confirm that the mapping is performed within the range of the appropriate phonon excitation number.

The boson expansion of the norm operator in Eqs. (\ref{eq:znexp}) seemingly suggests that the convergence appears to be better than the case without CAA-C, which is, however, caused by the overestimation of the effect of the Pauli exclusion principle. The neglected terms in CAA-C contribute to weakening the effect of the Pauli exclusion principle improperly enhanced by CAA-C. 
The boson expansions without CAA-C incorporate more exactly the effect of the Pauli exclusion principle than those using CAA-C does.

It has been claimed that the closed-algebra approximation, used to obtain the recurrence formula of the multi-phonon matrix element, is a good one and that the boson expansion of Hamiltonian converges well because the multi-phonon matrix elements rapidly become zero as the increased number of phonons due to the effects of the Pauli exclusion principle \cite{HJJ76}.
This claim is not correct in the first place.
The boson expansions converge well not because of the strong effect of the Pauli exclusion principle but because of its weakness.
The purpose of the limitation of the sort and the excitation number of the phonons is to weaken the effect of the Pauli principle.
It is for the same reason that the convergence of the expansion is improved and that the zero eigenvalue does not appear in the multi-phonon norm matrix up to a sufficient number of excitations when we select only the phonons with collective excitation modes. 
It is the weak effect of the Paul exclusion principle that makes the phonons with collective excitation modes behave like bosons and the boson expansion methods successful.
\color{black}

Once $\hat Z(N)$ is known, we can obtain $\overline{X_{\bar t}}(N)$ from the following recurrence formula derived from Eqs. (\ref{eq:barxbtdagn}),
\begin{equation}
\begin{array}{lll}
\overline{X_{\bar t'}}(N)&=&-\displaystyle\frac 12\sum_{t_1}\sum_{t'_1t'_2}Y(\bar t't'_1t'_2t_1)\hat Z(N)b_{t_1}^\dagger b_{t'_1}b_{t'_2}
\\
&&+\displaystyle\frac 14\sum_{t_1t_2}\sum_{t'_1t'_2t'_3}\sum_{\bar t}Y(\bar t't'_2t'_3\bar t)Y(\bar t t_1t_2\bar t'_1)b_{t_1}^\dagger b_{t_2}^\dagger b_{t'_1}\hat Z(N-1)b_{t'_2}b_{t'_3}
\\
&&+\displaystyle\frac 14\sum_{t_1t_2}\sum_{t'_1t'_2}\sum_{\bar t}\sum_{\bar t'_1}Y(\bar t't'_1t'_2\bar t)Y(\bar tt_1t_2\bar t'_1)b_{t_1}^\dagger b_{t_2}^\dagger\overline{X_{\bar t'_1}}(N-2)b_{t'_1}b_{t'_2},
\end{array}
\end{equation}
and find the solutions as
\begin{subequations}
\begin{equation}
\label{eq:olxbtN}
\overline{X_{\bar t'}}(N)=(X_{\bar t'})_L\mathcal{\hat Z}\hat1_B(N+1)=\hat1_B(N)(X_{\bar t'})_L\mathcal{\hat Z},
\end{equation}
\begin{equation}
 (X_{\bar t'})_L
= (X_{\bar t'})_L^{(2)}+(X_{\bar t'})_L^{(3)}+(X_{\bar t'})_L^{(4)}+O(\Gamma^5),
\end{equation}
\begin{equation}
\begin{array}{lll}
(X_{\bar t'})_L^{(2)}&=&-\displaystyle\frac 12\sum_{t_1}\sum_{t'_1t'_2}Y(\bar t't'_1t'_2t_1)b_{t_1}^\dagger b_{t'_1}b_{t'_2},
\\
(X_{\bar t'})_L^{(3)}&=&\displaystyle\frac 14\sum_{t_1t_2}\sum_{t'_1t'_2t'_3}\sum_{\bar t}Y(\bar t't'_2t'_3\bar t)Y(\bar tt_1t_2t'_1)b_{t_1}^\dagger b_{t_2}^\dagger b_{t'_1}b_{t'_2}b_{t'_3},
\\
(X_{\bar t'})_L^{(4)}&=&[\mathcal{\hat Z}^{(2)}, (X_{\bar t'})_L^{(2)}]
\\
&&-\displaystyle\frac18\sum_{t_1t_2t_3}\sum_{t'_1t'_2t'_3t'_4}\sum_{\bar t\bar t''}Y(\bar t't'_1t'_2\bar t)Y(\bar tt_1t_2\bar t'')Y(\bar t''t'_3t'_4t_3)b_{t_1}^\dagger b_{t_2}^\dagger b_{t_3}^\dagger b_{t'_1}b_{t'_2}b_{t'_3}b_{t'_4},
\end{array}
\end{equation}
\begin{equation}
\begin{array}{l}
[\mathcal{\hat Z}^{(2)}, (X_{\bar t'})_L^{(2)}]=-\displaystyle\frac 14\sum_{t_1}\sum_{t'_1t'_2}\sum_{tt'}Y(\bar t'tt't_1)Y(t'_1tt't'_2)b_{t_1}^\dagger b_{t'_1}b_{t'_2}
\\
\qquad-\displaystyle\frac 14\sum_{t_1t_2}\sum_{t'_1t'_2t'_3}\sum_t\left\{2Y(t_1t'_1t'_2t)Y(t\bar t't_2t'_3)-Y(\bar t't'_1t'_2t)Y(tt_1t_2t'_3)\right\}b_{t_1}^\dagger b_{t_2}^\dagger b_{t'_1}b_{t'_2}b_{t'_3}.
\end{array}
\end{equation}
\end{subequations}
From these, we obtain
\begin{equation}
\overline{X_{\bar t'}}=\sum_{N=0}^{N_{max}-1}\overline{X_{\bar t'}}(N)=(X_{\bar t'})_L\mathcal{\hat Z}\breve1_B=\breve 1_B^{(-1)}(X_{\bar t'})_L\mathcal{\hat Z}=\breve 1_B(X_{\bar t'})_L\mathcal{\hat Z}\breve 1_B,
\end{equation}
where
\begin{equation}
\breve 1_B^{(\Delta N)}=\sum_{N=0}^{N_{max}+\Delta N}\hat 1_B(N).
\end{equation}
$\breve 1_B^{(0)}=\breve 1_B$ and $\breve 1_B\breve 1_B^{(-1)}=\breve 1_B^{(-1)}\breve 1_B=\breve 1_B^{(-1)}$ hold.
Organizing the Hermitian conjugate of Eq. (\ref{eq:olxbtN}), we find
\begin{subequations}
\begin{equation}
\begin{array}{lll}
\overline{X_{\bar t}^\dagger}(N)&=&((X_{\bar t})_L\mathcal{\hat Z}\hat1_B(N+1))^\dagger
=\hat1_B(N+1)\mathcal{\hat Z}(X_{\bar t})_L^\dagger
\\
&=&\mathcal{\hat Z}(X_{\bar t})_L^\dagger\hat1_B(N)
=\left\{\mathcal{\hat Z}(X_{\bar t})_L^\dagger\mathcal{\hat Z}^{-1}\right\}\mathcal{\hat Z}\hat 1_B(N)
\\
&=&(X_{\bar t}^\dagger)_L\mathcal{\hat Z}\hat 1_B(N),
\end{array}
\end{equation}
where
\begin{equation}
(X_{\bar t}^\dagger)_L=\mathcal{\hat Z}(X_{\bar t})_L^\dagger\mathcal{\hat Z}^{-1},
\end{equation}
\begin{equation}
 (X_{\bar t}^\dagger)_L
=(X_{\bar t}^\dagger)_L^{(2)}+(X_{\bar t}^\dagger)_L^{(3)}+(X_{\bar t}^\dagger)_L^{(4)}+O(\Gamma^5),
\end{equation}
\begin{equation}
\begin{array}{lll}
(X_{\bar t}^\dagger)_L^{(2)}&=&((X_{\bar t})_L^{(2)})^\dagger,\quad (X_{\bar t}^\dagger)_L^{(3)}=((X_{\bar t})_L^{(3)})^\dagger,
\\
(X_{\bar t}^\dagger)_L^{(4)}&=&((X_{\bar t})_L^{(4)})^\dagger+[\mathcal{\hat Z}^{(2)}, ((X_{\bar t})_L^{(2)})^\dagger]
\\
&=& -\displaystyle\frac 18\sum_{t_1t_2t_3t_4}\sum_{t'_1t'_2t'_3}\sum_{\bar t'\bar t''}Y(\bar tt_1t_2\bar t')Y(\bar t't'_1t'_2\bar t'')Y(\bar t''t_3t_4t'_3)b_{t_1}^\dagger b_{t_2}^\dagger b_{t_3}^\dagger b_{t_4}^\dagger b_{t'_1} b_{t'_2} b_{t'_3},
\end{array}
\end{equation}
\end{subequations}
and obtain
\begin{equation}
\label{eq:barxbartdagg}
\overline{X_{\bar t}^\dagger}=\sum_{N=0}^{N_{max}-1}\overline{X_{\bar t}^\dagger}(N)=(X_{\bar t}^\dagger)_L\mathcal{\hat Z}\breve 1_B^{(-1)}=\breve1_B(X_{\bar t}^\dagger)_L\mathcal{\hat Z}=\breve 1_B(X_{\bar t}^\dagger)_L\mathcal{\hat Z}\breve 1_B.
\end{equation}
In this way, we can obtain $(X_{\bar t'})_L$ and $(X_{\bar t}^\dagger)_L$ as infinite expansions.

Specifically dealing with the terms up to $O(\Gamma^4)$, we found that $\hat Z(N)=\mathcal{\hat Z}\hat 1_B(N)=\hat 1_B(N)\mathcal{\hat Z}$, $\overline{X_{\bar t'}}(N)=\overline{X_{\bar t'}}\hat 1_B(N+1)=\hat 1_B(N)\overline{X_{\bar t'}}$, and $\overline{X_{\bar t}^\dagger}(N)=\hat 1_B(N+1)\overline{X_{\bar t}^\dagger}=\overline{X_{\bar t}^\dagger}\hat 1_B(N)$.
Oppositely, assuming that these hold for any $N$, and substituting them into Eq. (\ref{eq:barxbtdagn}) and Eq. (\ref{eq:zn}), we can find the relational expressions for $\mathcal{\hat Z}$, $\overline{X_{\bar t'}}$, and $\overline{X_{\bar t}^\dagger}$, and solve these for each order of magnitude, then we obtain the same results. This result suggests that the $N$ dependency of these operators generally holds. 

Applying the above results to Eqs. (\ref{eq:uoun}) and summing up $N$, we obtain
\begin{subequations}
\begin{equation}
\begin{array}{lll}
\overline{X_{t'}} &=&(X_{t'})_L\mathcal{\hat Z}\breve 1_B=\breve 1_B^{(-1)}(X_{t'})_L\mathcal{\hat Z}=\breve 1_B(X_{t'})_L\mathcal{\hat Z}\breve 1_B,
\\
(X_{t'})_L&=&(X_{t'})_D,
\end{array}
\end{equation}
\begin{equation}
\begin{array}{lll}
\overline{X_t^\dagger}&=&\breve 1_B(X_t^\dagger)_L\mathcal{\hat Z}=(X_t^\dagger )_L\mathcal{\hat Z}\breve 1_B^{(-1)}=\breve 1_B(X_t^\dagger )_L\mathcal{\hat Z}\breve 1_B,
\\
(X_t^\dagger)_L&=&(X_t^\dagger )_D+(X_t^\dagger )_{out}
\\
(X_t^\dagger )_{out}&=& -\displaystyle\frac 12\sum_{t_1t_2}\sum_{\bar t'_1}Y(tt_1t_2\bar t'_1)b_{t_1}^\dagger b_{t_2}^\dagger (X_{\bar t'_1})_L,
\end{array}
\end{equation}

\begin{equation}
\begin{array}{lll}
\label{eq:barbqn}
\overline{B_q}&=&(B_q)_L\mathcal{\hat Z}\breve 1_B=\breve 1_B(B_q)_L\mathcal{\hat Z}=\breve 1_B(B_q)_L\mathcal{\hat Z}\breve 1_B,
\\
 (B_q) _L&=&(B_q)_D+(B_q)_{out},
\\
(B_q)_{out}&=&\displaystyle\sum_t\sum_{\bar t'}\Gamma_q^{\bar t't}b_t^\dagger(X_{\bar t'})_L.
\end{array}
\end{equation}
\end{subequations}

While, from $B_q=B_{\bar q}^\dagger$, $\overline{B_q}=\overline{B_{\bar q}}^\dagger$ and we find another expression for $\overline{B_q}$ as
\begin{equation}
\begin{array}{lll}
\overline{B_q}&=&\breve 1_B\mathcal{\hat Z}(B_{\bar q})_L^\dagger =\mathcal{\hat Z}(B_{\bar q})_L^\dagger\breve 1_B=\breve 1_B\mathcal{\hat Z}(B_{\bar q})_L^\dagger\breve 1_B,
\\
 (B_{\bar q}) _L^\dagger&=&(B_q)_D+(B_{\bar q})_{out}^\dagger,
\\
(B_{\bar q})_{out}^\dagger&=&\displaystyle\sum_t\sum_{\bar t}\Gamma_q^{t'\bar t}(X_{\bar t})_L^\dagger b_{t'},
\end{array}
\end{equation}
where we use $(B_q)_D=(B_{\bar q})_D^\dagger$.
Using two types of expressions for $\overline{B_q}$, we obtain
\begin{equation}
\label{eq:bqdzcom}
\breve 1_B[(B_q)_D, \mathcal{\hat Z}]\breve 1_B=\breve 1_B\{\mathcal{\hat Z}(B_{\bar q})_{out}^\dagger-(B_q)_{out}\mathcal{\hat Z}\}\breve 1_B.
\end{equation}

From Eq. (\ref{eq:bmopzubar}) and Eq. (\ref{eq:normopresp}), we can express the mapping operator as
\begin{equation}
U_\xi=\mathcal{\hat Z}^{\xi-\frac 12}\bar U,
\end{equation}
and Eq. (\ref{eq:ximapzbar}) becomes as follows:
\begin{subequations}
\label{eq:ximapcalzbar}
\begin{equation}
\label{eq:ximapcalzbar1}
\vert \psi')_{\xi}= \mathcal{\hat Z}^{\xi-\frac 12}\overline{\vert\psi')},\qquad {}_{-\xi} (\psi\vert=\overline{(\psi\vert}\mathcal{\hat Z}^{-\xi-\frac 12},
\end{equation}
\begin{equation}
\label{eq:ximapcalzbar2}
(O_F)_{\xi}=\mathcal{\hat Z}^{\xi-\frac 12}\overline{O_F}\mathcal{\hat Z}^{-\xi-\frac 12},
\end{equation}
\end{subequations}
If $O_F$ is a phonon creation, a phonon annihilation, or a scattering operator, $\overline{O_F}=\breve 1_B(O_F)_L\mathcal{\hat Z}\breve 1_B$ holds. Therefore the mapped $O_F$ can be expressed as
\begin{equation}
\label{eq:ofxi}
(O_F)_\xi=\breve 1_B(O_F)_{B(\xi)}\breve 1_B;\quad (O_F)_{B(\xi)}=\mathcal{\hat Z}^{\xi-\frac 12}(O_F)_L\mathcal{\hat Z}^{-\xi+\frac 12},
\end{equation}
and 
\begin{equation}
{}_{-\xi}(\psi\vert(O_F)_\xi\vert\psi')_\xi={}_{-\xi}(\psi\vert(O_F)_{B(\xi)}\vert\psi')_\xi
\end{equation}
holds. Therefore we can regard $(O_F)_\xi$ as $(O_F)_{B(\xi)}$ in the physical subspace.
The boson expansions of $(O_F)_{B(\xi)}$ become infinite expansions for an arbitrary $\xi$ because those of $(O_F)_L$ become infinite expansions.

For $\xi\neq 0$,  the boson expansions become non-Hermitian types. In the case of $\xi=\frac12$,  $(O_F)_{\xi(\frac 12)}=(O_F)_L$ holds. 
Application of
\color{blue} CAA-C
\color{black} derives DBET.

For $\xi=0$, the boson expansions become the Hermitian type and can be derived using
\begin{equation}
\begin{array}{lll}
(O_F)_{B(0)}&=&\mathcal{\hat Z}^{-\frac 12}(O_F)_L\mathcal{\hat Z}^{\frac 12}
\\
&=&(O_F)_L+\frac 12[(O_F)_L, \mathcal{\hat Y}]
-\frac 38\mathcal{\hat Y}[(O_F)_L, \mathcal{\hat Y}]-\frac 18[(O_F)_L, \mathcal{\hat Y}]\mathcal{\hat Y}+O(\Gamma^6).
\end{array}
\end{equation}

The boson expansions of the phonon creation and annihilation operators and the scattering operators are as follows:
\begin{subequations}
\label{eq:xtg4}
\begin{equation}
(X_{t'})_{B(0)}=(X_{t'})_{B(0) in}+(X_{t'})_{B(0) out},
\end{equation}
\begin{equation}
(X_{t'})_{B(0) in}=b_{t'}+(X_{t'})_{B(0) in}^{(2)}+(X_{t'})_{B(0) in}^{(4)}+O(\Gamma^5),
\end{equation}
\begin{equation}
(X_{t'})_{B(0) in}^{(2)}=-\frac14\sum_{t_1}\sum_{t'_1t'_2}Y(t't'_1t'_2t_1)b_{t_1}^\dagger b_{t'_1}b_{t'_2},
\end{equation}
\begin{equation}
\begin{array}{r}
(X_{t'})_{B(0) in}^{(4)}=-\displaystyle\frac{1}{32}\sum_{t_1}\sum_{t'_1t'_2}\sum_{tt''}Y(t'tt''t'_1)Y(t'_2tt''t'_1)b_{t_1}^\dagger b_{t'_1}b_{t'_2}
\\
+\displaystyle\frac 1{96}\sum_{t_1t_2}\sum_{t'_1t'_2t'_3}\sum_t\{2Y(t'_1t_1t't)Y(tt'_2t'_3t_2)
\\
-5Y(t't'_1t'_2t)Y(tt_1t_2t'_3)\}b_{t_1}^\dagger b_{t_2}^\dagger b_{t'_1}b_{t'_2}b_{t'_3},
\end{array}
\end{equation}
\begin{equation}
(X_{t'})_{B(0)out}=(X_{t'})_{B(0)out}^{(3)}+(X_{t'})_{B(0)out}^{(4)}+O(\Gamma^5)
\end{equation}
\begin{equation}
\begin{array}{r}
(X_{t'})_{B(0)out}^{(3)}=\displaystyle\frac 1{24}\sum_{t_1t_2}\sum_{t'_1t'_2t'_3}\sum_{\bar t}\{2Y(t'_1t_1t'\bar t)
Y(\bar tt'_2t'_3t_2)
\\
+Y(t'_1t_1t_2\bar t)Y(\bar tt'_2t'_3t')\}b_{t_1}^\dagger b_{t_2}^\dagger b_{t'_1}b_{t'_2}b_{t'_3},
\end{array}
\end{equation}
\begin{equation}
\begin{array}{r}
(X_{t'})_{B(0)out}^{(4)}=-\displaystyle\frac 1{16}\sum_{t_1t_2t_3}\sum_{t'_1t'_2t'_3t'_4}\sum_{\bar t\bar t'}Y(t'_1t't_1\bar t)Y(\bar tt'_2t'_3\bar t')Y(\bar t't_2t_3t'_4)
\\
b_{t_1}^\dagger b_{t_2}^\dagger b_{t_3}^\dagger b_{t'_1}b_{t'_2}b_{t'_3}b_{t'_4}.
\end{array}
\end{equation}
\end{subequations}
\begin{subequations}
\label{eq:xtbarg4}
\begin{equation}
(X_{\bar t'})_{B(0)}=(X_{\bar t'})_{B(0)}^{(2)}+(X_{\bar t'})_{B(0)}^{(3)}+(X_{\bar t'})_{B(0)}^{(4)}+O(\Gamma^5) ,
\end{equation}
\begin{equation}
(X_{\bar t'})_{B(0)}^{(2)}=(X_{\bar t'})_L^{(2)},\quad (X_{\bar t'})_{B(0)}^{(3)}=(X_{\bar t'})_L^{(3)},
\end{equation}
\begin{equation}
(X_{\bar t'})_{B(0)}^{(4)}=(X_{\bar t'})_L^{(4)}-\frac 12[ \mathcal{\hat Z}^{(2), }(X_{\bar t'})_L^{(2)}].
\end{equation}
\end{subequations}
\begin{subequations}
\label{eq:bqg4}
\begin{equation}
\begin{array}{lll}
(B_q)_{B(0)}&=&(B_q)_L+\frac 12\mathcal{\hat Z}\{(B_{\bar q})_{out}{}^\dagger-(B_q)_{out}\}+O(\Gamma^5),
\\
&=&(B_q)_{B(0)in}+(B_q)_{B(0)out},
\end{array}
\end{equation}
\begin{equation}
(B_q)_{B(0)in}=(B_q)_D,
\end{equation}
\begin{equation}
(B_q)_{B(0)out}=(B_q)_{B(0)out}^{(2)}+(B_q)_{B(0)out}^{(3)}+(B_q)_{B(0)out}^{(4)}+O(\Gamma^5),
\end{equation}
\begin{equation}
(B_q)_{B(0)out}^{(k)}=\frac 12\{(B_q)_{out}^{(k)}+(B_{\bar q})_{out}^{(k)}{}^\dagger\}\quad (k=2,3),
\end{equation}
\begin{equation}
(B_q)_{B(0)out}^{(4)}=\frac 12\{(B_q)_{out}^{(4)}+(B_{\bar q})_{out}^{(4)}{}^\dagger\}+\frac 12\mathcal{\hat Z}^{(2)}\{(B_{\bar q})_{out}^{(2)}{}^\dagger-(B_q)_{out}^{(2)}\},
\end{equation}
\begin{equation}
\begin{array}{r}
\frac 12\mathcal{\hat Z}^{(2)}\{(B_{\bar q})_{out}^{(2)}{}^\dagger-(B_q)_{out}^{(2)}\}=\displaystyle\frac 18\sum_{t_1t_2}\sum_{t'_1t'_2}\sum_{tt'}\sum_{\bar t}\{\Gamma_q^{t'_1\bar t}Y(\bar ttt't'_2)-\Gamma_q^{\bar tt}Y(\bar tt'_1t'_2t')\}Y(tt_1t_2t')
\\
b_{t_1}^\dagger b_{t_2}^\dagger b_{t'_1}b_{t'_2}
\\
+\displaystyle\frac 18\sum_{t_1t_2t_3}\sum_{t'_1t'_2t'_3}\sum_{t}\sum_{\bar t}\{2\Gamma_q^{t'_1\bar t}Y(\bar tt_1tt'_2)-\Gamma_q^{\bar tt}Y(\bar tt'_1t'_2t_1)-\Gamma_q^{\bar tt_1}
Y(\bar tt'_1t'_2t)\}
\\
Y(tt_1t_2t'_3)b_{t_1}^\dagger b_{t_2}^\dagger b_{t_3}^\dagger b_{t'_1}b_{t'_2}b_{t'_3}
\\
+\displaystyle\frac1{16}\sum_{t_1t_2t_3t_4}\sum_{t'_1t'_2t'_3t'_4}\sum_{\bar t}\{\Gamma_q^{t'_1\bar t}Y(\bar tt_1t_2t'_2)-\Gamma_q^{\bar tt_1}Y(\bar tt'_1t'_2t_2)\}
Y(t'_4t_3t_4t'_3)
\\
b_{t_1}^\dagger b_{t_2}^\dagger b_{t_3}^\dagger b_{t_4}^\dagger b_{t'_1}b_{t'_2}b_{t'_3}b_{t'_4}.
\end{array}
\end{equation}
\end{subequations}
Here, we use Eq. (\ref{eq:bqdzcom}) to find $(B_q)_{B(0)}$. From Eqs. (\ref{eq:barbqn}), $(B_q)_{out}^{(k)}=\sum_t\sum_{\bar t'}\Gamma_q^{\bar t't}b_t^\dagger(X_{\bar t'})_L^{(k)}$.

The terms derived in NOLCEXP are only those of the order of magnitude up to $O(\Gamma^2)$ except for $(X_{\bar t'})_{B(0)in}$ among these, and NOLCEXP derives $(B_q)_{B(0)out}^{(2)}$ as higher-order than $O(\Gamma^2)$ \cite{KT76, KT83, SK88, SK91}.

The first term of $(X_{t'})_{B (0)in}^{(4)}$ has two $t$ -sums, which becomes large in case of inappropriate adoption of the phonon excitation modes. This term causes trouble with the convergence of the boson expansion when adopting all or too many phonon excitation modes. Therefore we should appropriately adopt the phonon excitation modes and the maximum phonon excitation number for convergence.

The phonon-truncation approximation
\color{blue}
that DBET adopts
\color{black}
excludes the unselected phonon operators, while the boson expansions without applying the closed-algebra approximations have these terms as higher-order terms, $(X_{\bar t'})_B{(\xi)}$, $(X_ {\bar t}^\dagger)_{B (\xi) }\sim O(\Gamma^2)$.
Applying \color{blue} CAA-C \color{black}, we obtain $(X_{\bar t'})_B{(\xi)}\approx 0 $, $(X_ {\bar t}^\dagger)_{B (\xi) }\approx 0$.
Different $\xi$ makes $ (B_q) _ {B (\xi)} $ have different expansions, while applying \color{blue} CAA-C\color{black}, we obtain $(B_q)_{B(\xi)} \approx (B_q) _D $ for all $\xi$, which result in the same as those of NOLCEXP in the case neglecting $\bar t$-sum, and  those of DBET. 
The order of magnitude of the neglected parts is $O(\Gamma^2)$.
In this way, applying \color{blue} CAA-C \color{black} in the case where the small parameter expansion holds brings the same results as the phonon-truncation approximation. 
The order of magnitude of the terms that \color{blue} CAA-C \color{black} neglects is $O(\Gamma^2)$, which is the same as the next-to-leading order terms of the boson expansions for the adopted phonons. 
\color{blue}
$\{t\}$ must be small enough for the small parameter expansion, and the application of CAA-C results in, in this case, the loss of the non-negligible contribution of the unselected phonon excitation modes.

NOLCEXP can not be achieved without no limitation of the phonon excitation number under the assumption that the norm matrices have no zero eigenvalues in any case \cite{KT83}.
It is stated that for this assumption to hold, it is necessary to map the subspace of the fermion space rather than the entire fermion space \cite{SK88}, which is, however, not enough for the case where the number of phonon excitations becomes infinite. 
No clear reason has been given to use this seemingly unnatural assumption.
We should make it clear whether we can justify it or not.

We can justify limiting the number of bosons after finding the boson expansions as the small parameter expansions without limiting the number of phonon excitations as follows:
Eq. (\ref{eq:ofxi}), which is an equation obtained on the assumption that the norm operator does not have zero eigenvalues within the range of phonon excitation numbers below $N_ {max}$, indicates that the limitation of the phonon excitation number is aggregated to $\breve1_B$. 
Therefore we can replace $(O_F)_\xi $ with $(O_F)_{B(\xi)}$ in the physical space. 
This replacement can be also virtually achieved as follows:
\begin{equation}
(O_F)_\xi\rightarrow (O_F)_{B(\xi)}\quad (N_{max}\rightarrow \infty),
\end{equation}
because
\begin{equation}
\breve 1_B\rightarrow\hat 1_B\quad (N_{max}\rightarrow \infty)
\end{equation}
holds.
It indicates that we can obtain the correct boson expansions by performing small parameter expansion under the artificial assumption that the eigenvalues of the norm operator do not become zero at any boson excitation number.  It is a matter of course that after obtaining those, we should work within the appropriate $N_{max}$ for the ideal boson state vector to be physical.

We can also justify this assumption within the framework of NOLCEXP. NOLCEXP uses the boson expansion of the projection operator to the boson vacuum:
\begin{equation}
\label{eq:0pro}
\vert 0)(0\vert =\hat 1_B+\sum_{N=1}^{\infty}\frac{1}{N!}b_{t_1}^\dagger\cdots b_{t_N}^\dagger b_{t_1}\cdots b_{t_N}.
\end{equation}
Since $\vert 0)(0\vert$  is expanded as normal-ordered, the limitation of the excitation number of bosons  permits the following cutoff:
\begin{equation}
\label{eq:0prolim}
\vert 0)(0\vert =\hat 1_B+\sum_{N=1}^{N_{max}}\frac{1}{N!}b_{t_1}^\dagger\cdots b_{t_N}^\dagger b_{t_1}\cdots b_{t_N},
\end{equation}
where $N_{max}$ is the maximum excitation number of bosons, which is also that of phonons.
The larger $N_ {max}$ becomes, it requires more expansion terms because it is not a small parameter expansion. 
Even if $ N_ {max} $ is set so that the multi-phonon norm matrices have no zero eigenvalues, it would not be so small.
To obtain the small parameter expansions using norm matrices expansions, we should take up all terms of Eq. (\ref{eq:0prolim}) when boson excitations are allowed up to $ N_ {max} $.
Therefore, it is not enough to confirm that the small parameter expansions hold where the number of boson excitations is small. On the other hand, the part exceeding $ N_ {max} $ does not affect the result obtained up to $ N_ {max} $ due to the normal-ordered expansions, which enables us to impose any assumption for the norm matrices. 
It justifies the infinite expansions under seemingly unnatural assumptions of NOLCEXP.  Nothing to say, the expansions obtained in this way are valid only within the range of phonon excitation numbers in which zero eigenvalues do not appear in the norm matrices of the multi-phonon state vectors.

Although the problem of setting seemingly unnatural assumptions has been settled in this way, NOLCEXP has still difficulties.
Utilizing the boson expansion of the projection operator onto the boson vacuum and the expansion of the multi-phonon norm matrices requires a great deal of effort to show that the expansions have fast convergence as the normal-ordered linked-cluster type even with CAA-C.
On the other hand, the norm operator method does not require the boson expansion of the projection operator onto the boson vacuum, and can easily obtain the boson expansion of the norm operator, etc., all of which become of linked-cluster type.  Eq. (\ref{eq:ofxi}) indicates that the boson expansion obtained as a result of combining them is not in the normal order. 
Making them into normal order, contractions between the boson creation and annihilation operators occur to generate only new $t$-sums, and they again become of linked-cluster type without changing their order of magnitude of the boson expansions.
Eqs. (\ref{eq:xtg4}) ,(\ref{eq:xtbarg4}), and (\ref{eq:bqg4}) shows these results.
The norm operator method is a new method for deriving linked-cluster expansions, from which we can derive {\it normal-ordered linked-cluster expansions} without the closed-algebra approximation.

A more serious difficulty with NOLCEXP is the use of CAA-C to claim its legitimacy.
Marshalek claimed that the boson expansions where the phonon operators become infinite normal-ordered expansions and the scattering operators become finite have no physical subspace, and he called them chimerical \cite{Ma80a, Ma80b}.
NOLCEXP has used CAA-C to obtain the normal-ordered expansions for phonon operators, and the finite expansions of the scattering operators.
There is no success of CAA-C, however, without making small parameter expansions impossible even from the stage of two-phonon excitations, which means that NOLCEXP does not hold in this case.
The small parameter expansions do not hold because the CAA-C makes the norm matrices have 0 eigenvalues at the stage of 2-phonon excitation.
It also makes the ideal boson state vectors contain spurious components.
We can not use the ideal boson state vectors as physical in this case.
NOLCEXP has appealed that the ideal boson state vectors can be used as physical and that the linked-cluster expansion brings excellent convergence.
The success of CAA-C spoils both of the appealing points.
DBET, which needs CAA-C to make its expansions finite, does not suffer the convergence problem but can not use the ideal boson state vectors as physical as far as it claims the superiority of its finite expansions.
It does not mean that there is no physical subspace because the projection operator onto the physical subspace exists, which can remove the spurious components.
It is not, however, practical and undermines the advantages of the boson expansion methods.
Marshalek's claim that no physical subspace exists should be replaced by that the physical subspace can not be spanned by the ideal boson state vectors.
Therefore it should be claimed that CAA-C makes not only NOLCEXP   become {\it chimerical} but also DBET. 
Although Marshalek has withdrawn the claim of chimerical \cite{KM91}, we have made it clear, in a different approach, that CAA-C makes boson expansions become {\it chimerical}.

Both the refutation of NOLCEXP for chimerical and the claim for the superiority of the finite expansions of DBET are not correct because they are based on the improper use of the closed-algebra approximation.

NOLCEXP and DBET have used the ideal boson state vectors actually, which are not allowed under the successful application of CAA-C.
Therefore we should pick up the neglected terms by CAA-C, which makes it possible to include the effect of the Pauli exclusion principle more accurately.
This correction makes all types of boson expansions become infinite ones.
DBET can no longer claim the superiority of its finite expansions.

\color{black}
The boson expansion method, which does not adopt
\color{blue}
CAA-C, not only avoids those defects that NOLCEXP and DBET have but also
\color{black}
enables us to incorporate non-collective excitation modes in a way not possible with the conventional practical boson expansion methods.
The only way to incorporate non-collective phonon excitation modes using
\color{blue}
CAA-C
\color{black}
is to introduce bosons that bear the excitation modes corresponding to those of the phonons.
The small parameter expansions require the restriction of the number of sorts of the non-collective modes for taking $\{t\}$ small enough.
On the other hand, without
\color{blue}
CAA-C,
\color{black}
we can incorporate the contributions of all
\color{blue}
or enough
\color{black}
sorts of non-collective modes into the coefficients of boson expansions without introducing the bosons having the non-collective modes.
\color{blue}
This method incorporating the contribution of the non-collective modes into the coefficient is much easier than that of the diagonalization introducing the non-collective modes as boson excitations \cite{TTT87}.
\color{black}
The boson expansions consisting of only the bosons having the collective modes have good convergence under \color{blue} CAA-C \color{black} \cite{KT76, KT83}.
\color{blue}Although there is an overestimation of the effect of the Pauli exclusion principle by CAA-C, \color{black}
the estimation for the order of magnitude for the boson expansions without closed-algebra approximation\color{blue}, which treats the Paluli exclusion principle more accurately, \color{black} suggests that the boson expansions with only bosons having the collective modes converge well also in this case.
\color{blue}

It is the kinematical point of view that determines whether we should adopt the closed-algebra approximation or not.
The boson expansions with proper or no use of the closed-algebra approximation enable us to include enough or all modes neglected by the improper application of the closed-algebra approximation, which results in treating the commutation relations among the phonons and the scattering operator more accurately, that is, treating the effect of the Pauli exclusion principle more accurately.
We call this method of inclusion of the neglected non-collective modes {\it kinematical inclusion}.

It has been, so far, only the dynamical point of view that determines which non-collective excitation mode should be selected.
In the application to the real system,  the collective phonon excitation modes are selected among the quadrupole modes considering the dynamics of the system \cite{KT76, TTT87, SK91}.
How to adopt the non-collective excitation modes is divided into some cases, adopting some of the quadrupole modes \cite{TTT87}, all of them \cite{KT76}, and all those with spin parity from $0^+$ to $4^+$ \cite{SK91}. The reason for the selection in the last case is that they are directly coupled to the collective excitation modes with the number of excitation 1 or 2.
The contribution of these modes is introduced as bosons.
We call this method of inclusion of the non-collective modes {\it dynamical inclusion}.

{\it Kinematical inclusion} is involved in how to incorporate the effect of the Pauli exclusion, while {\it dynamical inclusion} is involved in selecting phonons that should be boson-expanded related to the dynamics of the system.
We should couple these two ways of inclusion of the non-collective excitation modes and investigate them, for example, as follows: Take only the collective excitation modes as $\{t\} $, adopt the non-collective modes, having only $2^+$, from $0^+$ to $4^+$, and more widely, as the components of $\{\bar t\}$ and compare them.
In addition, include the non-collective modes having only $2^+$ , from $0^+$ to $4^+$ into $\{t\} $ and take other non-collective modes into $\{\bar t\}$.
\color{black}

Finally, we deal with the product of operators.
Let $O_F$ and $O'_F$ be the phonon creation, annihilation operators, or scattering operators, respectively, we can derive the boson expansions of their product as 
\begin{equation}
\label{eq:ofof'exp}
(O_FO'_F)_{B(\xi)}=\mathcal{\hat Z}^{\xi-\frac 12}\overline{O_FO'_F}\mathcal{\hat Z}^{-\xi-\frac 12}.
\end{equation}
If $\overline{O_FO'_F}=\breve 1_B(O_F)_L\breve 1_B(O'_F)_L\mathcal{\hat Z}\breve 1_B$ holds,  we obtain
\begin{equation}
\label{eq:ofof'exppr}
(O_FO'_F)_{B(\xi)}=(O_F)_{B(\xi)}(O'_F)_{B(\xi)},
\end{equation}
and if Eq. (\ref{eq:ximapzbar22aprox}) holds,
\begin{equation}
\label{eq:ofof'expprapp}
(O_FO'_F)_{B(\xi)}\approx (O_F)_{B(\xi)}(O'_F)_{B(\xi)}.
\end{equation}
In the case that Eq. (\ref{eq:ofof'exppr}) and Eq. (\ref{eq:ofof'expprapp}) hold, it is sufficient to 
obtain only the boson expansions of the basic fermion pair operators.
Conventional practical boson expansion methods have used the approximation of Eq. (\ref{eq:ofof'expprapp}) as a matter of course.
Eq. (\ref{eq:ofof'exp}) makes it possible to judge whether this approximation is good or bad.
We present $\overline{O_FO'_F}$ in the appendix.

\subsection{On the boson expansions when the \color{blue} conventional \color{black} closed-algebra approximation holds}
\color{blue}
We have clarified that CAA-C does not hold in the small parameter expansions. Here, we try to show what kind of boson expansion we can obtain when CAA-C holds.
\color{black}

When \color{blue} CAA-C \color{black} is applicable, the norm matrix of the two-phonon state vectors approximately has zero eigenvalues.  Therefore, small parameter expansions break down. 
For example, we can express $\mathcal{\hat Z}_{in}^{(4)}$ of Eq. (\ref{eq:zn4}), obtained under small parameter expansions, as
\begin{equation}
\label{eq:z4in}
\begin{array}{lll}
\mathcal{\hat Z}_{in}^{(4)}&=&-\displaystyle\frac1{16}\sum_{t_1t_2}\sum_{t'_1t'_2}\sum_{tt'}Y(t_1tt't_2 )Y(t'_1ttt'_2)b_{t_1}^\dagger b_{t_2}^\dagger b_{t'_1}b_{t'_2}
\\
&&-\displaystyle\frac 1{24}\sum_{t_1t_2t_3}\sum_{t'_1t'_2t'_3}\sum_tY(t'_1t_1t_2t)Y(tt'_2t'_3t_3)b_{t_1}^\dagger b_{t_2}^\dagger b_{t_3}^\dagger b_{t'_1}b_{t'_2}b_{t'_3}
\\
&&+\displaystyle\frac 12 (\mathcal{\hat Z}^{(2)})^2.
\end{array}
\end{equation}
the double $t$-sums appear in the first term on the right side.
As mentioned earlier, this term does not become $O(\Gamma^4)$ with adopting all kinds of phonon excitation modes or when \color{blue} CAA-C \color{black} holds, and, in addition, it makes the small parameter expansions impossible.
Therefore we can not use the expressions after Eqs. (\ref{eq:znrec}), which hold only under small parameter expansions. Instead, we should apply \color{blue} CAA-C \color{black} turning back to Eq. (\ref{eq:uoun}). 
We obtain
\begin{subequations}
\label{eq:uouncaa}
\begin{equation}
\overline{X_{t'}}(N)=(X_{t'})_D\hat Z(N+1)\qquad (N\geq 0),
\end{equation}

\begin{equation}
\label{eq:barxtdagncaa}
\begin{array}{lll}
\overline{X_t^\dagger}(0)&=&(X_t^\dagger )_D\hat Z(0),
\\
\overline{X_t^\dagger}(N)&\approx&(X_t^\dagger )_D\hat Z(N)\qquad (N\geq 1),
\end{array}
\end{equation}

\begin{equation}
\begin{array}{lll}
\label{eq:barbqncaa}
\overline{B_q}(0)&=&0,
\\
\overline{B_q}(N) &\approx&(B_q)_D\hat Z(N)\qquad (N\ge 1),
\end{array}
\end{equation}
\end{subequations}
\begin{equation}
\label{eq:barxbtdagncaa}
\begin{array}{lll}
\overline{X_{\bar t}^\dagger}(0)&=&0,
\\
\overline{X_{\bar t}^\dagger}(N)
&\approx&0\qquad (N\ge 1).
\end{array}
\end{equation}
Therefore, when $O_F$ is one of $X_{t'} $, $ X_t^\dagger$, or $B_q$, we obtain
\begin{equation}
(O_F)_\xi\approx\hat Z^{\xi-\frac 12}(O_F)_D\hat Z^{-\xi+\frac 12},
\end{equation}
\begin{equation}
\label{eq:oo'caapprox}
(O_FO'_F)_\xi\approx\hat Z^{\xi-\frac 12}(O_F)_D(O'_F)_D\hat Z^{-\xi+\frac 12}.
\end{equation}
The latter holds because $\overline{O_FO'_F}\approx (O_F)_D(O'_F)_D\hat Z$ from Eq. (\ref{eq:ubarop}).

We also obtain
\begin{equation}
\label{eq:zncaapprox}
\hat Z(N)\approx\frac 1N\sum_t(X_t^\dagger)_D\hat Z(N-1)b_t=\frac1{N!}\sum_{t_1\cdots t_N}(X_{t_1}^\dagger)_D\cdots(X_{t_N}^\dagger)_D\vert 0)(0\vert b_{t_1}\cdots b_{t_N},
\end{equation}
and we can express the norm operator as
\begin{equation}
\hat Z\approx\vert 0)(0\vert+\sum_{N=1}^{N_{max}}\frac1{N!}\sum_{t_1\cdots t_N}(X_{t_1}^\dagger)_D\cdots(X_{t_N}^\dagger)_D\vert 0)(0\vert b_{t_1}\cdots b_{t_N}.
\end{equation}
We can not, however, expand it as a small parameter expansion.
For $\xi=\frac 12$, the norm operator does not appear in the mapped fermion operators, and we can obtain the boson expansions as follows:
\begin{equation}
\label{eq:oaaprox12}
(O_F)_{\frac 12}\approx\hat T_B(O_F)_D\hat T_B,
\end{equation}
\begin{equation}
\label{eq:oo'caapprox12}
(O_FO'_F)_{\frac 12}\approx\hat T_B(O_F)_D(O'_F)_D\hat T_B,
\end{equation}
which indicates that we obtain the finite expansions of DBET if $\hat T_B=\breve 1_B$ holds, which is not the case as mentioned before.

Finally, we make mention of $(B_q)_\xi$. From Eq. (\ref{eq:barbqncaa}), $[(B_q)_D, \hat Z(N)]\approx 0$ holds, then
\begin{equation}
\label{eq:bqzncom}
[(B_q)_D, \hat Z]\approx 0.
\end{equation}
Hence
\begin{equation}
(B_q)_\xi\approx (B_q)_D\hat T_B=\hat T_B(B_q)_D=\hat T_B(B_q)_D\hat T_B
\end{equation}
holds for any $\xi$.

\section{Comparison to boson-fermion expansion theory}
By replacing the collective modes $\{c\}$ and the non-collective modes $\{n\}$ to $\{t\}$ and $\{\bar t\}$, respectively and suppressing the fermion excitations, we can obtain the boson expansions from the boson-fermion expansions \cite{TM90, TM91}. 
The Hermitian-type boson expansions obtained here agree with those obtained from the boson-fermion expansion theory adopting the proper transformation \cite{TKM94}.

We have already compared the boson part of the boson-fermion expansion theory with the terms of NOLCEXP \cite{TKM94}, where we have derived order by order their terms discarded in  \color{blue} CAA-C \color{black} referring to the methods of the derivation of NOLCEXP. 
They coincide with the results of the systematic derivation here.

Further comparison requires the derivation of higher-order terms in the boson-fermion expansion theory.

\section{Summary}
We have proposed a new boson expansion theory, \color{blue} the norm operator method, \color{black} without using the closed-algebra approximation indispensable for their formulations as an extension of the conventional practical boson expansion methods that have tried to elucidate nuclear collective motion.

We have introduced a mapping operator that limits the phonon excitation number in addition to the phonon excitation modes, which have made it possible to perform small parameter expansions without the unnatural assumption. By introducing the norm operator into the mapping operator, we can perform boson expansions easier, which has enabled the boson expansions without the closed-algebra approximation.
Any types of boson expansion method become infinite small parameter expansions. We have obtained higher-order terms and those neglected under the closed-algebra approximation. 
\color{blue}The boson expansion of the norm operator not only facilitates the boson expansions but also makes it possible to confirm that the mapping is performed within the range of the appropriate phonon excitation number.
\color{black}

 We have shown that the boson approximation is nothing but the boson mapping with the maximum excitation number of phonons 1, which indicates that the boson approximation has been established as the boson mapping with the maximum excitation number of phonons 1.

\color{blue}
We have investigated the two-phonon norm matrix composed of all the phonon excitation modes in detail and revealed the mechanism, which is essential for the boson expansion methods, of how we should construct the fermion subspace to be mapped from the whole fermion space.
\color{black}

\color{blue}
Utilizing the two-phonon norm matrix, 
\color{black}
we have proposed a rule for estimating the order of magnitude for the boson expansions as small parameter expansions that do not use the closed-algebra approximation and given the condition that the small parameter expansions hold. 
We have also found the cause of the failure of the small parameter expansion and obtained the boson expansions in which the expansion terms that cause the failure do not appear.

\color{blue}
The closed-algebra approximation, which closes the commutation relations among the selected phonons, compels the following two-phonon matrices to have the nesting structure: Any two-phonon norm submatrix whose excitation modes are closed among them has the same structure as the two-phonon norm matrix with all excitation modes, which makes it necessarily include a zero eigenvalue.
As a result, the closed-algebra approximation makes it impossible to remove all the non-collective modes having no correspondence to bosons proper way for the boson expansion methods.
\color{black}
The closed-algebra approximation
\color{blue}
that closes the commutation relations among the phonons to be boson expanded
is applicable only when the small parameter expansions become impossible and the ideal boson state vectors become unphysical.

The conventional boson expansion methods have applied the closed-algebra approximation for the small parameter expansions improper way actually, which results in neglecting the terms from the next-to-leading order of magnitude and improperly strengthens the effect of the Pauli principle.

It is the kinematical point of view that determines whether we should adopt the closed-algebra approximation or not.
The proper application of the closed-algebra approximation needs enough phonon excitation modes that have no correspondence to the boson excitation modes.

The norm operator method permits us to include the contribution of all the phonon excitation modes that have no correspondence to the boson excitation modes and the proper application of the closed-algebra approximation enables us to remove the modes not effective from them, which results in treating the commutation relations among the phonons and the scattering operator more accurately, that is, treating the effect of the Pauli exclusion principle more accurately.

For the success of the small parameter expansions, we should select, as those to be the boson excitations, not only the phonon excitation modes having enough high collectivity but also should not adopt too many phonon excitation modes.
And the excitation modes having more collectivity needs more excitation modes not treated as boson excitations, which makes the small parameter expansions more successful and the convergence of the boson expansion better.

Both the refutation of NOLCEXP for chimerical and the claim for the superiority of the finite expansions of DBET have no validity because they are based on the improper use of the closed-algebra approximation.
NOLCEXP can not refute the claim that the normal-ordered boson expansions with those for the scattering operators as finite expansions are chimerical because NOLCEXP can not derive the scattering operators as the finite expansions without making the small parameter expansions impossible. 
\color{black}
DBET, which use the non-Hermite type of mapping and derives the finite expansions not only for the scattering operators but also for phonon operators, can not claim its finite expansions superiority \color{blue}
without making the ideal boson state vectors include the spurious components, which is also the case for deriving the finiteness of the scattering operator of NOLCEXP.
\color{black}

We have also mentioned the boson-fermion expansion theory. The Hermitian-type boson expansions obtained in this article agree with those obtained from the boson-fermion expansion theory adopting the proper transformation.

\color{blue}
The norm operator method is a method that allows the closed-algebra approximation not to be used or to be used appropriately, enables us to obtain the boson expansion easier, and reproduces the fermion subspace onto the boson subspace more faithfully than the conventional practical methods.
\color{black}

\appendix
\section{Formulas of the product of the pair operators}
We denote $B_q$, $X_{t''}$, or $X_{\bar t'}$ as $O_F$. the following equations hold:
\begin{subequations}
\begin{equation}
\overline{X_{t'}O_F}=\breve 1_B(X_{t'})_LB(O_F)_L\mathcal{\hat Z}\breve 1_B,\end{equation}
\begin{equation}
\overline{O_F^\dagger X_t^\dagger}=\breve 1_B(O_F^\dagger)_L(X_t^\dagger)_L\mathcal{\hat Z}\breve 1_B.
\end{equation}
\end{subequations}

\begin{equation}
\overline{X_t^\dagger X_{t'}}=\breve 1_B\left\{(X_t^\dagger)_D(X_{t'})_D-\frac 12\sum_{t_1t_2}\sum_{\bar t'}Y(tt_1t_2\bar t'_1)b_{t_1}^\dagger b_{t_2}^\dagger (X_{t'})_L(X_{\bar t'_1})_L\right\}\mathcal{\hat Z}\breve 1_B.
\end{equation}

\begin{subequations}
\begin{equation}
\overline{X_{\bar t}^\dagger X_{\bar t'}^\dagger}=\breve 1_B(X_{\bar t}^\dagger)_L(X_{\bar t'}^\dagger)_L\mathcal{\hat Z}\breve 1_B+O(\Gamma^5),
\end{equation}

\begin{equation}
\overline{X_{\bar t} X_{\bar t'}}=\breve 1_B(X_{\bar t})_L(X_{\bar t'})_L\mathcal{\hat Z}\breve 1_B+O(\Gamma^5).
\end{equation}
\end{subequations}

\begin{equation}
\overline{X_{\bar t}^\dagger X_{\bar t'}}=\breve 1_B(X_{\bar t}^\dagger)_L(X_{\bar t'})_L\mathcal{\hat Z}\breve 1_B+O(\Gamma^5).
\end{equation}

\begin{subequations}
\begin{equation}
\overline{ B_qX_{\mu'}}=(B_q)_D\overline{X_{\mu'}}+\sum_{t}\sum_{\bar t'}\Gamma_q^{\bar t't}b_t^\dagger\overline{X_{\bar t'}X_{\mu'}},
\end{equation}
\begin{equation}
\overline{X_{\mu}^\dagger B_q}=\overline{X_{\mu}^\dagger}(B_q)_D+\sum_{t'}\sum_{\bar t}\Gamma_q^{t'\bar t}\overline{X_{\mu}^\dagger X_{\bar t}^\dagger}b_{t'}.
\end{equation}

\end{subequations}

\end{document}